\documentclass[11pt,oneside,letterpaper]{article}
\usepackage{amssymb}
\usepackage{amsmath,bm}
\usepackage[dvips]{graphicx}
\usepackage{setspace}
\usepackage{fancyhdr}
\usepackage[dvipsnames]{xcolor}
\usepackage{ifpdf}
\usepackage{graphicx}
\usepackage{rotating}
\usepackage{comment}
\usepackage[margin=2.5cm]{geometry}
\usepackage[colorlinks=true, linkcolor  = blue,urlcolor=blue,citecolor=violet]{hyperref}
\usepackage{soul}
\usepackage{relsize}
\usepackage{tikz}
\usepackage{float}
\usepackage{empheq}
\usepackage{bbold}
\usepackage{cite}
\usepackage{etoolbox}
\patchcmd{\thebibliography}{\section*{\refname}}{}{}{}

\newcommand{\bi}{\begin{itemize}}
\newcommand{\ei}{\end{itemize}}
\newcommand{\bea}{\begin{eqnarray}}
\newcommand{\eea}{\end{eqnarray}}
\newcommand{\be}{\begin{equation}}
\newcommand{\ee}{\end{equation}}

\newcommand{\dd}{\mathrm{d}}
\newcommand{\im}{\mathrm{Im}}
\newcommand{\re}{\mathrm{Re}}
\numberwithin{equation}{section}
\allowdisplaybreaks  

\begin{document}



\onehalfspacing

\begin{center}

~
\vskip4mm
{{\huge {
\quad The two-sphere partition function in two-dimensional quantum gravity
 }
  }}
\vskip5mm

\vskip2mm

\vskip10mm


Dionysios Anninos$^1$, Teresa Bautista$^1$, and ~Beatrix M\"uhlmann$^2$ \\ 

\end{center}
\vskip4mm
\begin{center}
{
\footnotesize
{$^1$Department of Mathematics, King's College London, Strand, London WC2R 2LS, UK \newline\newline
$^2$Institute for Theoretical Physics and $\Delta$ Institute for Theoretical Physics, University of Amsterdam, \\Science Park 904, 1098 XH Amsterdam, The Netherlands\\
}}
\end{center}
\begin{center}
{\textsf{\footnotesize{
dionysios.anninos@kcl.ac.uk, teresa.bautista@kcl.ac.uk, 
beatrix@ias.edu}} } 
\end{center}
\vskip5mm

\vspace{4mm}
 
\vspace*{0.6cm}

\vspace*{1.5cm}
\begin{abstract}
\noindent
We study the Euclidean path integral of two-dimensional quantum gravity with positive cosmological constant coupled to conformal matter with large and positive central charge. The problem is considered in a semiclassical expansion about a round two-sphere saddle. We work in the Weyl gauge whereby the computation reduces to that for a (timelike) Liouville theory. We present results up to two-loops, including a discussion of contributions stemming from the gauge fixing procedure. We exhibit cancelations of ultraviolet divergences and provide a path integral computation of the central charge for timelike Liouville theory. Combining our analysis with insights from the DOZZ formula we are led to a proposal for an all orders result for the two-dimensional gravitational partition function on the  two-sphere.

\end{abstract}

\newpage
\setcounter{page}{1}
\pagenumbering{arabic}

\tableofcontents

\section{Introduction}

In this work we explore theories of two-dimensional quantum gravity coupled  to conformal matter. Our approach rhymes with that of \cite{Zamolodvarphikov:1982vx,Menotti:2004xz,Harlow:2011ny} in that we will rely heavily on the existence of a semiclassical limit. Although gravity in two-dimensions is a rather simple theory, when coupled to conformal matter it exhibits various features in common with a theory of gravity in a four-dimensional world. Particularly, the theory can have real semiclassical saddle point solutions \cite{Polchinski:1989fn}, and it moreoever exhibits a direct analogue of the unbounded conformal mode rendering the Euclidean gravitation action an infamously subtle affair. The advantage of the two-dimensional approach, as has been explored and emphasised often in the past and also more recently, is the remarkable control one has over the gravitational path integral \cite{Polyakov:1981rd,Distler:1988jt,David:1988hj}. 

The specific class of gravitational theories we consider are ones equipped with a positive cosmological constant $\Lambda$. We will couple these theories to  two-dimensional conformal matter endowed with a large and positive central charge $c_{\text{m}}$, thereby introducing a large number of field theoretic degrees of freedom. Examples of two-dimensional conformal field theories with $c_{\text{m}}\gg 1$ include tensor product theories, and holographic CFTs.\footnote{In this case, coupling to two-dimensional quantum gravity can be viewed as rendering the boundary metric of the dual AdS$_3$ geometry dynamical \cite{Compere:2008us}.} As we shall soon see, the large value of $c_{\text{m}}$ is crucially tied to the existence of a semiclassical picture. The technical object we will focus on will be the gravitational path integral over the matter and gravitational fields on an $S^2$ topology. We will offer a physical motivation for this shortly. The main goal of our work is to provide the systematics for a perturbative quantum loop expansion about the round two-sphere saddle in the large $c_{\text{m}}$ limit. We apply this to compute the saddle point approximation and the first few orders in the quantum loop expansion of the two-sphere partition function of interest. We work in the Weyl gauge whereby the gravitational path integral takes the form of a (timelike) Liouville theory \cite{Distler:1988jt,David:1988hj}. From this perspective, our approach offers a path integral perspective for the semiclassical expansion of the timelike DOZZ formula proposed and explored in \cite{Harlow:2011ny,Zamolodchikov:2005fy,Kostov:2005kk,Giribet:2011zx,McElgin:2007ak}. 

Before embarking on our analysis we find it instructive to provide a physical derivation and motivation for the saddle point solutions of the gravitational theories under consideration. 

\subsection*{Two-dimensional gravity coupled to conformal matter}


The theory we will focus on consists of a two-dimensional matter CFT of central charge $c_{\text{m}}$ coupled to two-dimensional gravity. The Euclidean path integral of interest is given by
\begin{equation}\label{zgrav0}
\mathcal{Z}_{\text{grav}}[\Lambda] = \sum_{h=0}^\infty e^{\vartheta \chi_h}\int [\mathcal{D} g_{ij} ] \, e^{- \Lambda \int_{\Sigma_h} \dd^2 x \sqrt{g} } \times Z^{(h)}_{\text{CFT}}[g_{ij}]~,
\end{equation}
where we take $\Lambda >0$, $\chi_h$ is the Euler character of the genus $h$ surface $\Sigma_h$, and $Z^{(h)}_{\text{CFT}}[g_{ij}]$ is the matter CFT partition function. The positivity of $\Lambda$ suppresses large area configurations. Unless otherwise stated we will, for the most part, restrict our attention to $S^2$ topology. 

Although the pure gravity theory has no classical solutions in two-dimensions due to the topological nature of the Einstein term, upon coupling to a matter theory the effective gravitational action including the contribution from $Z^{(h)}_{\text{CFT}}[g_{ij}]$ does \cite{Polchinski:1989fn}. If we further require that the saddle be semiclassical, we must additionally take the central charge $c_{\text{m}}$ to be large. Exploiting that two-dimensional gravity is invariant under the two-dimensional diffeomorphism group we can fix the metric, at least within a small neighbourhood, to the Weyl gauge
\begin{equation}\label{weylgauge}
ds^2 =  e^{2  \varphi(x)} \tilde{g}_{ij} \dd x^i \dd x^j~.
\end{equation}
Here $\varphi(x)$ is a real valued function encoding the Weyl factor of the physical metric. For vanishing genus we can further choose $\tilde{g}_{ij}$ as the round metric on $S^2$ with area $4\pi$, namely 
\begin{equation}
{d\tilde{s}^2} = \tilde{g}_{ij} \dd x^i \dd x^j =  \dd\theta^2 + \sin^2\theta \dd\phi^2~,
\end{equation}
with $\theta \in (0,\pi)$ and $\phi \sim \phi + 2\pi$. On an $S^2$ topology, and more specifically on an $S^2$ geometry, the form of  $Z_{\text{CFT}}^{(0)}[g_{ij}]$ is known \cite{Polyakov:1981rd,Zamolodchikov:2001dz}. Up to local ultraviolet divergences that can be absorbed into the bare gravitational couplings $\vartheta$ and $\Lambda$ the form of $Z_{\text{CFT}}^{(0)}[g_{ij}]$ is fixed by the conformal anomaly. Concretely, for genus zero
\begin{equation}\label{CFTS2}
Z_{\text{CFT}}^{(0)}[g_{ij}] = \mathcal{N} e^{-S_{\text{anomaly}}[\varphi]}~, 
\end{equation}
where $\mathcal{N}$ is a normalisation constant, and $S_{\text{anomaly}}[\varphi]$ is\footnote{More covariantly, the anomaly action is given by the Polyakov action whose form is 
\begin{equation}
S_{\text{anomaly}} = -\frac{c_{\text{m}}}{96\pi} \int \dd^2 x \sqrt{g} R \nabla_g^{-1} R~.
\end{equation}
Although the action is non-local, the degree of non-locality seems permissible in the context of two-dimensional gravity. Here it stems from integrating out the massless matter content of the two-dimensional conformal field theory. 
We further note that if we allow $\tilde{g}_{ij}$ to vary, $\mathcal{N}$ should be viewed as a functional of $\tilde{g}_{ij}$ and the  gravitational action acquires the structure of a WZW model (see for instance \cite{Ferrari:2011rk,Nink:2015lmq}).
}
\begin{equation}
S_{\text{anomaly}}[\varphi] = - \frac{c_{\text{m}}}{48\pi} \int \dd^2 x \sqrt{\tilde{g}} \left( 2 \tilde{g}^{i j} \partial_i \varphi \partial_j \varphi  + {2} \tilde{R} \varphi \right)~.
\end{equation}
Consequently, in the semiclassical limit, the resulting action governing the Weyl mode is given by 
\begin{equation}\label{seff}
S_{\text{eff}}[\varphi]  = -\frac{c_{\text{m}}}{48\pi} \int \dd^2 x \sqrt{\tilde{g}} \left( 2 \tilde{g}^{i j} \partial_i \varphi \partial_j \varphi  + {2} \tilde{R} \varphi  \right) + \Lambda \int \dd^2 x \sqrt{\tilde{g}} e^{2 \varphi}~.
\end{equation}
Interestingly, much like what happens for the conformal mode in higher dimensional Euclidean gravity, $S_{\text{eff}}$ has a wrong sign kinetic term for $\varphi$. Being semiclassical, $S_{\text{eff}}$ ignores contributions from the $\mathfrak{b}\mathfrak{c}$-ghost sector as well as the gravitational path integration measure. The reason is that these contributions, though present, will be subleading at large $c_{\text{m}}$. They will constitute much of what follows in the later sections. 

The saddle point equations stemming from (\ref{seff}) admit a constant $\varphi$ saddle with
\begin{equation}\label{grsaddle}
\varphi_* = \frac{1}{2} \log \frac{c_{\text{m}}}{24\pi \Lambda}~.
\end{equation} 
The saddle point solution corresponds to a round two-sphere of area  $\rho_* = c_{\text{m}}/6\Lambda$. To leading order, the vanishing genus expression reads:
\begin{equation}\label{leadingZ}
\log \mathcal{Z}_{\text{grav}}^{(0)}[\Lambda] = 2 \vartheta +    \frac{c_{\text{m}}}{6} \log  {\rho}_* + \text{const}~, 
\end{equation}
where the constant is $\Lambda$ independent, but will generally depend on $c_{\text{m}}$.  For $\vartheta \gg 1$, higher genus corrections to (\ref{zgrav0}) are exponentially small as compared to the genus zero contribution. 
Thus, two-dimensional gravity coupled to a CFT with large positive central charge admits a classical (Euclidean) dS$_2$ solution 
\begin{equation}\label{ds2metrics}
{ds^2} = \frac{\rho_*}{4\pi} \left(\dd\theta^2 + \sin^2\theta \dd\phi^2\right)~,
\end{equation}
with $\theta \in (0,\pi)$ and $\phi \sim \phi+2\pi$, which is the round two-sphere. Upon Lorentzian continuation $\phi \to i t$,  we are led to 
\begin{equation}
 {ds^2} = \frac{\rho_*}{4\pi} \left(\dd\theta^2 - \sin^2\theta \dd t^2\right)~,
\end{equation}
with $t\in\mathbb{R}$. This is the static patch of dS$_2$ with the de Sitter horizon residing at $\theta=0$ and $\theta=\pi$. 

Intuitively, the physical reason the matter CFT can support a positive curvature space with positive gravitational vacuum energy density $\Lambda$ is due to the negative static patch energy $E_{\text{s.p.}} = -c_{\text{m}}/{12\pi \ell}$ (see for instance section 7 of \cite{Anninos:2020hfj}) associated to a matter CFT quantised on a fixed dS$_2$ background with de Sitter length $\ell$ (corresponding to a round two-sphere of area $\rho_{\text{dS}} = 4\pi \ell^2$ upon Wick rotating to Euclidean signature). General relativity on a compact spatial slice with no boundaries enforces that the total energy vanishes, leading to the expression
\begin{equation}
\Lambda -\frac{c_{\text{m}}}{24\pi \ell^2} = 0~,
\end{equation}
which is solved by $\ell = \sqrt{c_{\text{m}}/(24\pi \Lambda)}$, precisely corresponding to the saddle (\ref{grsaddle}).

According to Gibbons and Hawking \cite{Gibbons:1976ue,Gibbons:1977mu} the gravitational path integral $\log\mathcal{Z}_{\mathrm{grav}}[\Lambda]$ in (\ref{zgrav0}) computes the entropy of the dS$_2$ horizon. What is missing as compared to higher-dimensions is a term related to the area in Planck units, since there is no Planck length (or horizon area) in two-dimensions. Instead, the term $2\vartheta$ in (\ref{zgrav0}), stemming from the two-dimensional Einstein term, might be viewed as the lower-dimensional analogue of the tree-level horizon entropy. The logarithmic term (\ref{leadingZ}) is the entanglement entropy of the CFT degrees of freedom across the dS$_2$ horizon \cite{Holzhey:1994we, Calabrese:2004eu}. Indeed, for a CFT on the static patch of dS$_2$ with de Sitter length $\ell$ the entanglement entropy across the dS$_2$ horizon is \cite{Casini:2011kv}
\begin{equation}\label{Sent}
S_{\text{ent}} = \frac{c_{\text{m}}}{6}\log {\rho_{\text{dS}}} + \text{const}~,
\end{equation}
where the constant is $\rho_{\text{dS}}$ independent while generally depending on $c_{\text{m}}$. It should be clear then how to compare (\ref{Sent}) with the logarithmic term in (\ref{leadingZ}).

\subsection*{Outline}

In the following sections we will explore the semiclassical expansion about the saddle (\ref{grsaddle}) being more precise about the subleading corrections. We will discuss more detailed properties of $\mathcal{Z}_{\mathrm{grav}}[\Lambda]$ (\ref{zgrav0}) at the quantum level perturbatively near the classical saddle. Corrections away from the standard CFT expression (\ref{Sent}) are due to the interaction with gravity. In this sense, the analysis provides us a window into the effects of gravity on entanglement entropy in two-dimensions. In section \ref{weylsec} we discuss how to fix the Weyl gauge, including a careful treatment of the residual $PSL(2,\mathbb{C})$ gauge symmetries of $\mathcal{Z}^{(0)}_{\mathrm{grav}}[\Lambda]$. In section \ref{sctl} we compute various perturbation corrections to $\mathcal{Z}_{\text{grav}}^{(0)}[\Lambda]$, culminating in the two-loop expression (\ref{eq:SL_final_pathInt}). In section \ref{dozzsec} we compare our results to a prediction of the exact sphere partition function stemming from an analytic continuation (\ref{tLZs2}) of the DOZZ formula. We conclude with an outlook in section \ref{outlook}. The appendices provide important details leading to the results presented in the main text.

\section{The Weyl gauge}\label{weylsec}

In order to proceed, we will consider the problem in the Weyl gauge (\ref{weylgauge}). In doing so, we employ a hypothesis of Distler-Kawai \cite{Distler:1988jt} and David \cite{David:1988hj} that fixes the path-integration measure over the Weyl mode $\varphi$. In this section, we consider the problem on a genus zero surface and provide a proper treatment of the residual gauge freedom. 

\subsection{Path integral in the Weyl gauge $\&$ timelike Liouville.} 

In complex coordinates the two-dimensional metric in the Weyl gauge can be expressed as
\begin{equation}\label{Weylcomplex}
ds^2 = e^{2b\varphi(z,\bar{z})} d\tilde{s}^2~,
\end{equation}
where our choice of fiducial metric takes the Fubini-Study form
\begin{equation}\label{FSmetric}
d\tilde{s}^2 = \frac{4 \upsilon \text{d} {z} \text{d} \bar{z} }{\left(1+ z \bar{z}\right)^2}~,
\end{equation}
with area $4\pi \upsilon$.
We have included the parameter  $b$ for future convenience. It will also be convenient to work with spherical coordinates
\begin{equation}
z = e^{i \phi} \, \tan \frac{\theta}{2}~, \quad\quad \bar{z} = e^{-i \phi} \, \tan \frac{\theta}{2}~,
\end{equation}
such that the fiducial metric is given by 
\begin{equation}\label{spherical}
d\tilde{s}^2 = \upsilon\left( \dd\theta^2 + \sin^2\theta \dd\phi^2 \right)~,
\end{equation}
with $\theta \in (0,\pi)$ and $\phi \sim \phi+2\pi$, capturing the two-sphere with area $4\pi \upsilon$.

In order to properly fix the Weyl gauge (\ref{Weylcomplex}) we must introduce Fadeev-Popov $\mathfrak{b}\mathfrak{c}$-ghost fields. Upon integrating out the matter and $\mathfrak{b}\mathfrak{c}$-ghost fields, our resulting gravitational path integral on a genus zero surface is given by 
\begin{equation}\label{g0Z}
\mathcal{Z}_{\text{grav}}^{(0)}[\Lambda] =  e^{2\vartheta} \times \frac{\mathcal{A}}{\text{vol}_{PSL(2,\mathbb{C})}} \times  \upsilon^{(c_{\text{m}}-26)/6} \times \int [\mathcal{D}\varphi] \,  e^{-S_L[\varphi]}~. 
\end{equation}
We have collected all dependence on the Weyl mode $\varphi$, including the contributions from the conformal anomalies of the matter and $\mathfrak{b}\mathfrak{c}$-ghost theories, into a single action which we have called $S_L[\varphi]$. We have also used the general expression (\ref{Sent}) to extract the $\upsilon$ dependence from the matter CFT and $\mathfrak{b}\mathfrak{c}$-ghost theory \cite{Zamolodchikov:2001dz}, where the constant $\mathcal{A}$ is $Z^{(0)}_{\text{CFT}}[\tilde{g}_{ij}] \times Z^{(0)}_{\mathfrak{b}\mathfrak{c}}[\tilde{g}_{ij}]$ with $\upsilon=1$.

According to the hypothesis of Distler-Kawai \cite{Distler:1988jt} and David \cite{David:1988hj}, this action is given by the Liouville action \cite{Polyakov:1981rd}
\begin{equation}\label{SL}
S_{L}[\varphi] = \frac{1}{4\pi} \int\dd^2x\sqrt{\tilde{g}} \left( \tilde{g}^{ij} \partial_i \varphi \partial_j \varphi +  Q \tilde{R} \varphi + 4\pi \Lambda e^{2b \varphi}  \right)~,
\end{equation}
with $[\mathcal{D}\varphi]$ being the standard flat measure on the space of fields $\varphi$.\footnote{In \cite{DHoker:1990dfh, DHoker:1990prw} a proof of this hypothesis for the spacelike case with $Q^2 \ge 2$ is presented.} Moreover, $Q = b+1/b$ and the Liouville central charge is $c_L = 1+6 Q^2$. Consistency of the theory, viewed as a theory of gravity coupled to conformal matter, requires $c_L - 26 + c_{\text{m}} = 0$. In terms of the matter central charge, 
\begin{equation}
Q = \sqrt{\frac{25-c_{\text{m}}}{6}}~, \quad\quad b =  \frac{\sqrt{25-c_{\text{m}}}-\sqrt{1-c_{\text{m}}}}{2 \sqrt{6}}~,
\end{equation} 
where the positive root for $Q$ is a choice we can make given the redundancy $Q \to - Q~$, $b\to -b$, and $\varphi \to - \varphi$. On the other hand, $b$ is chosen to be the solution admitting a semiclassical limit at large $c_{\text{m}}$. It would be interesting to explore the other solution of $b$ also.

For $c_{\text{m}} > 25$, $Q$ and $b$ are pure imaginary. To render the action real, we will consider the field $\varphi$ to live on a purely imaginary contour. The resulting theory is known as timelike Liouville theory \cite{Polyakov:1981rd}. It is convenient to parameterise this theory as follows:
\begin{equation}\label{eq:StL}
S_{tL}[\varphi]= \frac{1}{4\pi} \int \dd^2x\sqrt{\tilde{g}} \left( -\tilde{g}^{ij} \partial_i \varphi \partial_j \varphi -  q \tilde{R} \varphi + 4\pi \Lambda e^{2\beta\varphi}  \right)~,
\end{equation}
where $\beta =  i b$, $q = - i Q$ ({or similarly $\beta= -i b$ and $q = i Q$}) and $\varphi$ has been continued to reside along the real axis. 
The reality properties of $S_{tL}$ more closely resemble those of the effective action (\ref{seff}) obtained from our gravitational considerations. The cost of having a real action is a kinetic term which is unbounded from below. This is the two-dimensional analogue of the unbounded conformal mode in higher-dimensional theories of Euclidean gravity. 

In section \ref{sctl} we study the semiclassical expansion of $S_{tL}$. But before doing so, we must first discuss the residual gauge symmetries that remain upon fixing the Weyl gauge.

\subsection{Residual gauge symmetries \& further gauge fixing.}\label{resgauge}

The condition (\ref{Weylcomplex}) does not fully fix the gauge and ensues further redundancies. For instance, the transformation
\begin{equation}\label{redundant}
2b\varphi(z,\bar{z}) \to 2b\varphi(z,\bar{z})  - \sigma(z,\bar{z}) ~, \quad\quad \tilde{g}_{ij}(z,\bar{z})  \to e^{\sigma(z,\bar{z}) } \tilde{g}_{ij}(z,\bar{z})~,
\end{equation}
is a redundancy of the parametrisation. As such, the resulting theory must be invariant under the above redundancy. Locally, conformal Killing vectors of $\tilde{g}_{ij}$ are given by holomorphic maps
\begin{equation}\label{holomorphic}
z \to f(z)~, \quad \bar{z} \to \overline{f(z)}~.
\end{equation}
The above maps do not affect the form of the Weyl gauge (\ref{Weylcomplex}) since they can be reabsorbed in a shift of $\varphi(z,\bar{z})$. Of the space of holomorphic maps, we must select the subset of (\ref{holomorphic}) which are normalisable under the norm on the space of diffeomorphisms $\omega^i(z,\bar{z})$ \cite{Zamolodvarphikov:1982vx,Polyakov:1981rd} 
\begin{equation}\label{norm_diffeos}
ds_\omega^2 = \int  \text{d} {z} \text{d} \bar{z}  \sqrt{g} g_{ij} \omega^i \omega^j~.
\end{equation}
This leaves a $PSL(2,\mathbb{C})$ subgroup of normalisable residual diffeomorphisms. 
Explicitly, these are the maps
\begin{equation}
f(z) =  \frac{a z + b}{c z + d}~, \quad\quad a d - b c = 1~,
\end{equation}
with $a$, $b$, $c$, and $d$ in $\mathbb{C}$, and their anti-holomorphic counterpart $\overline{f(z)}$.
At the infinitesimal level, the transformations are given by  linearly expanding near the identity $a = d= 1$ and $c = b =0$:
\begin{equation}\label{linearz}
z \to z + \delta b +  \left(\delta a-\delta d \right) z - \delta c \, z^2~,
\end{equation}
where we must further impose $\delta a = -\delta d$ to preserve $a d- b c=1$ to linear order. The maximally compact $SO(3)$ group of $PSL(2,\mathbb{C})$ is given  by restricting 
\begin{equation}
b \bar{b} + d \bar{d} = 1~, \quad\quad a \bar{b} + c \bar{d}  = 0~, \quad\quad a \bar{a} + c \bar{c} = 1~.
\end{equation}
We can parameterise this by
\begin{equation}
b= \sin\rho e^{i\alpha}~, \quad d = \cos\rho e^{i\gamma}~, \quad  a= e^{i\delta} \bar{d}~,\quad c = -e^{-i\delta} \bar{b}~,
\end{equation}
with $\alpha \in (0,2\pi)$, $\gamma \in (0,2\pi)$, and $\rho \in (0,\pi/2)$. Recalling that $a d - b c=1$ sets $\delta = 0$ such that we are left with three real parameters $\{\alpha,\rho,\gamma\}$ which are the Hopf coordinates of the $SO(3)$ group manifold with geometry given by the round $S^3$  
\begin{equation}\label{hopf}
ds^2 =  \dd\rho^2 + \dd\alpha^2 \sin^2\rho + \dd\gamma^2 \cos^2\rho~.
\end{equation}
The identity is $\gamma=\rho=0$ such that the infinitesimal transformations (\ref{linearz}) reduce to
\begin{equation}\label{SO3linear}
\delta b = -\delta \bar{c} = e^{i \alpha} \delta \rho~, \quad\quad \delta d = -\delta a = i \delta \gamma~.
\end{equation}
The remaining three deformations in (\ref{linearz}) generate the remaining $PSL(2,\mathbb{C})$ transformations. 




The presence of these residual transformations justifies why we must divide the partition function (\ref{g0Z}) by the volume of $PSL(2,\mathbb{C})$ \cite{Zamolodchikov:1995aa}
\begin{equation}\label{sl2cmeasure}
\text{vol}_{PSL(2,\mathbb{C})} = 2 \int_{\mathbb{C}^4} \text{d}^2 a \, \text{d}^2 b \, \text{d}^2 c \, \text{d}^2 \, d \times \delta\left(a d - b c - 1 \right)~.
\end{equation}
This volume must be treated with care as it is infinite. The residual $PSL(2,\mathbb{C})$ symmetry acts non-trivially on the Liouville field $\varphi(z,\bar{z})$. 
%
We can study this in the semiclassical limit where $\beta \approx 1/q \ll 1$. For the sake of simplicity, we introduce $\Omega(z,\bar{z})$ satisfying
\begin{equation}
d\tilde{s}^2 = \frac{4 \upsilon \text{d} {z} \text{d} \bar{z} }{\left(1+ z \bar{z}\right)^2} \equiv e^{2\Omega(z,\bar{z})}  \dd z \dd\bar{z}~.
\end{equation}
It follows that 
\begin{equation}
\tilde{R} = -8\times e^{-2\Omega(z,\bar{z})} \partial_z \partial_{\bar{z}} \Omega(z,\bar{z})~,
\end{equation}
and the Liouville action (\ref{SL}) reads
\begin{equation}
S_{tL}[\varphi] = \frac{1}{4\pi} \int \dd z \dd\bar{z} \left(2 \,\varphi  \partial_z \partial_{\tilde{z}} \varphi - q \, \sqrt{\tilde{g}}   \tilde{R} \varphi + 4\pi \Lambda  \sqrt{\tilde{g}}  e^{2\beta \varphi} \right)~. 
\end{equation}
As shown in appendix \ref{FPapp}, it is relatively straightforward to check that in the semiclassical limit the transformation
\begin{equation}\label{phiT}
\varphi(z,\bar{z}) \to \varphi(f(z),\overline{f(z)}) + \frac{q}{2} \log f'(z) + \frac{q}{2} \log \overline{f'(z)} + q \left(\Omega(f(z),\overline{f(z)} ) - \Omega(z,\bar{z}) \right)~,
\end{equation}
leaves $S_{tL}$ invariant.\footnote{Beyond this limit the classical Liouville action is no longer invariant, rather it is the quantum theory that must be invariant under (\ref{phiT}).}
Given the invariance of  $S_{tL}$, and assuming it persists at the quantum level, the path-integral over the Liouville field $\varphi$ will produce a term proportional to the volume $PSL(2,\mathbb{C})$. Thus, $\text{vol}_{PSL(2,\mathbb{C})}$ appears in both the numerator and denominator of the gravitational path integral (\ref{g0Z}) in the Weyl gauge. 

To fix this we must resort to further gauge-fixing. We can expand the Liouville field in a complete basis of real spherical harmonics
\begin{equation}\label{phiexp}
\varphi(\Omega) = \sum_{l,m} \varphi_{lm} Y_{lm}(\Omega)~,
\end{equation} 
with $\Omega$ a point on the round metric on $S^2$ with unit area, i.e. (\ref{spherical}) with $\upsilon=1$, and the $\varphi_{lm}$ are real valued. We have normalised the $Y_{lm}(\Omega)$ as
\begin{equation}
\int \dd\Omega Y_{lm}(\Omega)Y_{l' m'}(\Omega) = \delta_{l l'} \delta_{m m'}~.
\end{equation}
Further conventions and properties for the spherical harmonics are given in appendix \ref{Yapp}. 

Under an isometric map $\Omega \to \Omega'(\Omega)$ the $Y_{lm}(\Omega)$ map to linear combinations of $Y_{lm}(\Omega)$ with the same $l$. 
{This is because for each $l$, $Y_{lm}(\Omega)$ with $m\in [-l,l]$ furnish irreducible representations of $SO(3)$.} As a gauge-fixing condition we follow \cite{Distler:1988jt} and impose that $\varphi_{1m} = 0$ with $m=\{-1,0,1\}$.\footnote{One might wonder whether this gauge condition is permissible and whether it suffers from Gribov ambiguities \cite{Gribov:1977wm}. However, as pointed out in  \cite{Distler:1988jt}, in the semiclassical limit where we study small field fluctuations the issue is tamed. For instance, the sign of the Fadeev-Popov determinant (\ref{FPexact}) remains unchanged in the semiclassical $q\rightarrow \infty$ limit.} The reason this is a good gauge fixing procedure, as delineated in appendix \ref{FP}, is that the variation of $\delta \varphi$ under the three non-compact generators of $PSL(2,\mathbb{C})$ is precisely equal to the $l = 1$ modes. A different gauge fixing procedure is discussed in \cite{Maltz:2012zs}. This condition will remain unchanged under the action of the $SO(3)$ subgroup of $PSL(2,\mathbb{C})$, fixing three of the six parameters of {$PSL(2,\mathbb{C})$}. Infinitesimally, these are given by (\ref{linearz}) where we must now deform in directions {$\alpha_n$} that are outside of (\ref{SO3linear}). A definition of $\alpha_n$ and the gauge-fixing procedure is given in appendix \ref{FP}. 
 Given the general transformation (\ref{phiT}), the Fadeev-Popov determinant for the gauge choice $\varphi_{1m} = 0$ is
\begin{equation}
\Delta_{\text{FP}} = \det \frac{\delta \varphi_{1 m} }{\delta \alpha_n}~, \quad\quad m \in \{ -1,0,1\}~, 
\end{equation}
and $n$ ranges over the three non-$SO(3)$ directions of $PSL(2,\mathbb{C})$. $SO(3)$ invariance fixes the form of the above to be independent of the $\varphi_{1m}$ and ensures that $\Delta_{\text{FP}}$ takes the following structure
\begin{multline}\label{FPexact}
\Delta_{\text{FP}} = a_0 q^3 + a_1 q\sum_{m=-2}^2 \varphi^2_{2m} + a_2 \Big(\varphi_{2,0}^3+\frac{3}{2}\varphi_{2,0}(\varphi_{2,1}^2+\varphi_{2,-1}^2)+\frac{3}{2}\sqrt{3}\varphi_{2,2}(\varphi_{2,1}^2-\varphi_{2,-1}^2)\cr
+3\sqrt{3}\varphi_{2,1}\varphi_{2,-1}\varphi_{2,-2}-3\varphi_{2,0}(\varphi_{2,-2}^2+\varphi_{2,2}^2)\Big)~.
\end{multline}
As we delineate in appendix \ref{FP}, 
\begin{equation}\label{a0a1a2}
a_0\equiv -\frac{16}{3\sqrt{3}}\pi^{3/2}~,\quad a_1\equiv \frac{12}{5}\sqrt{3\pi}~,\quad a_2\equiv \frac{12}{5}\sqrt{\frac{3}{5}}~.
\end{equation}
At this stage, all but the $SO(3)$ isometry group of the original two-dimensional diffeomorphisms has been gauge fixed. Since this is a compact group, we can just divide out its volume explicitly. 
The volume of $SO(3)$, as computed from the measure (\ref{hopf}), is given by
\begin{equation}
\text{vol}_{SO(3)} = {2\pi^2}~.
\end{equation}



\section{Timelike Liouville, semiclassically}\label{sctl}

We are now at a position to make sense of the genus zero timelike Liouville path integral
\begin{equation}\label{ZL}
\mathcal{Z}_{tL}[\Lambda] =  \frac{1}{\text{vol}_{PSL(2,\mathbb{C})}} \times \int [\mathcal{D}\varphi ] e^{-S_{tL}[\varphi]}~.
\end{equation}
in a semiclassical expansion about its real saddle.
Our large parameter is taken to be $1/\beta$, {with  $\beta \to 0^+$}, such that the remaining parameters are to be understood in terms of a large $1/\beta$ expansion. We recall the exact relations
\begin{equation}\label{bexp}
q = \frac{1}{\beta}- \beta~, \quad\quad c_{\text{m}}   =\frac{6}{\beta^2} +13+6\beta^2~.
\end{equation}


\subsection{Semiclassical saddle and small fluctuations}

The classical equations of the Liouville action (\ref{eq:StL}) are given by
\begin{equation}\label{eoms}
-2\tilde{\nabla}^2 \varphi = 8\pi \beta\Lambda e^{2\beta \varphi} -\frac{2}{\upsilon}q~,
\end{equation}
where $-\tilde{\nabla}^2$ is the Laplacian with respect to $\tilde{g}_{ij}$. The equations (\ref{eoms}) admit a constant and real solution given by
\begin{equation}\label{saddle}
\varphi_{*}= \frac{1}{2\beta}\log\left(\frac{q}{4\pi\upsilon\Lambda\beta}\right)~.
\end{equation}
Due to the $PSL(2,\mathbb{C})$ invariance of the Liouville action, in addition to the above solution there is a continuous family of solutions to (\ref{eoms}) related to (\ref{saddle}) by $PSL(2,\mathbb{C})$ transformations. Upon fixing the gauge of the residual gauge symmetries (as discussed in section \ref{resgauge}) we collapse the continuous solution space down to the constant saddle  (\ref{saddle}). As emphasised in \cite{Harlow:2011ny}, in addition to the real saddle point solution, (\ref{eoms}) also admit complex solutions $\varphi_* + \pi i n /\beta$ with $n \in \mathbb{Z}$. In this section we will focus on the semiclassical expansion about the real saddle point solution.  

The saddle point approximation to the path-integral (\ref{ZL}) leads to
\begin{equation}\label{ZLclas}
\mathcal{Z}_{\text{saddle}}[\Lambda] = \frac{1}{\text{vol}_{SO(3)}} \times  \left(\frac{q }{4\pi e \Lambda \upsilon \beta } \right)^{\frac{q}{\beta}} \approx  \left(\frac{1}{\Lambda \upsilon \beta^2}\right)^{\frac{1}{\beta^2}}~.
\end{equation}
Comparing to (\ref{leadingZ}) for $\upsilon=1$ and recalling (\ref{bexp}), we see that the two agree to leading order in the semiclassical limit. It is also interesting to note that (\ref{ZLclas}) bears some resemblance to the critical exponent type expressions encountered in spacelike Liouville theory \cite{Knizhnik:1988ak,Anninos:2020ccj,Bilal:2014mla}.


Expanding the Liouville field $\varphi = \varphi_* + \delta\varphi$ in (\ref{eq:StL}) about the saddle (\ref{saddle}) we find the action governing fluctuations 
\begin{equation}\label{Spert}
S_{\text{pert}}[\delta\varphi] = \frac{1}{4\pi} \int \dd^2 x \sqrt{\tilde{g}} \left( -\tilde{g}^{ij} \partial_i \delta \varphi  \partial_j \delta \varphi  + \frac{2}{\upsilon}(1-\beta^2) \delta \varphi^2    + \frac{4}{3\upsilon}\beta \delta \varphi^3 + \frac{2}{3\upsilon}\beta^2 \delta \varphi^4 \ldots \right)~,
\end{equation}
where we have kept terms up to order $\beta^2$. Notice that the small correction in the mass is of the same order as the coefficient of the quartic interaction. We are interested in computing the path integral over the fluctuation field $\delta \varphi$
\begin{equation}\label{Zpert}
\mathcal{Z}_{\text{pert}}[\beta] = \int [\mathcal{D} \delta\varphi ] \times  \Delta_{\text{FP}}\times \prod_{m=\{-1,0,1\}} \delta(\delta\varphi_{1m}) \times e^{-S_{\text{pert}}[\delta\varphi]}~,
\end{equation}
in a large $1/\beta$ expansion. Interestingly, {since $\sqrt{\tilde{g}}$ yields a factor of $\upsilon$,} the fluctuation action is independent of $\Lambda$ and $\upsilon$ to all orders. But it will depend non-trivially on $\beta$, and it is this dependence we will proceed to explore. The path integration contour for (\ref{Zpert}) is over real values of the fluctuation field $\delta\varphi$. As such,  the wrong sign kinetic term renders the path integral ill-defined. A natural proposal, which we pursue here, is to rotate the contour of $\delta \varphi$ to pure imaginary values while keeping $\beta$ fixed. This is the two-dimensional version of the proposal of Gibbons, Hawking, and Perry \cite{Gibbons:1978ac} for dealing with the unbounded conformal mode in Euclidean gravity. In rotating $\delta\varphi \to -i \delta\varphi$ the kinetic term becomes bounded, up to a zero mode which is Gaussian unsuppressed. To deal with this we will Wick rotate it back, as discussed in \cite{Polchinski:1988ua}. As we will soon see, the Wick rotation of the zero mode will produce an overall phase in $\mathcal{Z}_{tL}[\Lambda]$. On the other hand, though the continued perturbative theory has several imaginary couplings, we will argue that aside from the overall phase, the corrections are all real.   

One final remark is in order, before computing some contributions is in order. From (\ref{Spert}) it looks like the fluctuation action is strongly coupled for some modes. This can be seen by expanding $\delta \varphi$ in spherical harmonics. The three $l=1$ harmonics have an {\it almost} zero mode at the Gaussian level which is lifted at order $\sim \beta^2$. Fortunately, these almost zero modes are eliminated by the gauge-fixing condition for the residual gauge freedom. 


\subsection{One-loop contribution}

The leading contribution beyond the saddle-point approximation stems from the Gaussian fluctuations. The quadratic action, after continuing $\delta \varphi \to -i \delta \varphi$, is given by
\begin{equation}\label{Spert2}
S_{\text{pert}}^{(2)}[\delta \varphi ]  = \frac{1}{4\pi} \int \dd^2x \sqrt{\tilde{g}} \left(\tilde{g}^{ij} \partial_i \delta \varphi \partial_j \delta \varphi - \frac{2}{\upsilon}(1-\beta^2) \delta \varphi^2  \right)~.
\end{equation}
We note that although the kinetic term is positive, the mass-squared of the fluctuation is negative.\footnote{It may also be worth noting that the value of the tachyonic mass squared in the strict $\beta\to 0^+$ limit is $-2$ (in units of $\upsilon$) which plays a special role in the representation theory of dS$_2$ \cite{Bros:2010wa} as it furnishes the discrete series. In four-dimensional de Sitter, the discrete series is associated to gauge fields. In our case $\varphi$ is a piece of the two-dimensional metric, itself a gauge field.} 

We now expand $\delta \varphi$ in a complete and orthonormal basis of real spherical harmonics, as in (\ref{phiexp}). In terms of these, the measure of path integration is taken to be
\begin{equation}
[\mathcal{D}\delta\varphi] = \prod_{l,m} \left(\frac{\Lambda_{\mathrm{uv}}\upsilon}{\pi}\right)^{\frac{1}{2}}{\text{d}  \delta \varphi_{lm}}~,
\end{equation}
such that
\begin{equation}
1 = \int [\mathcal{D}\delta \varphi] e^{- \Lambda_{\mathrm{uv}} \int \dd^2 x \sqrt{\tilde{g}} \, \delta\varphi(x)^2}~.
\end{equation}
To render the measure local with respect to $\tilde{g}_{ij}$ we need to introduce an ultraviolet scale $\Lambda_{\text{uv}}$ with units of inverse area. The eigenvalues and degeneracies of the spherical Laplacian $-\tilde{\nabla}^2$ on (\ref{spherical}) are
\begin{equation}\label{eigenvalues}
\lambda_l = \frac{1}{\upsilon} \, l(l+1)~, \quad\quad d_l = 2l+1~, \quad\quad l = 0,1,\ldots~.
\end{equation}
For the case at hand, the constant mode fluctuation must be treated separately since it leads to an unsuppressed Gaussian. 
After the dust settles, we find that up to the Gaussian contribution 
\begin{equation}\label{ZL2}
\mathcal{Z}_{tL}[\Lambda] = \mathcal{Z}_{\text{saddle}}[\Lambda] \times \mathcal{Z}^{(2)}_{\text{pert}}[\beta]~,
\end{equation}
where the Gaussian one-loop contribution, including the Fadeev-Popov gauge fixing term at the quadratic level in the fields with (\ref{a0a1a2}), is given by
\begin{equation}\label{Zpert22}
\mathcal{Z}^{(2)}_{\text{pert}}[\beta] \equiv \pm i a_0 q^3\left(\frac{2\pi\upsilon\Lambda_{\mathrm{uv}}}{\beta q}\right)^{\frac{1}{2}}  \left({\frac{\upsilon\Lambda_{\mathrm{uv}}}{\pi}}\right)^{\frac{3}{2}}\left(\frac{4\pi\upsilon\Lambda_{\mathrm{uv}}}{6-2\beta q -4\pi\frac{a_1}{a_0q^2}}\right)^{\frac{5}{2}}\prod^\infty_{l=3}\left(\frac{4\pi\upsilon\Lambda_{\mathrm{uv}}}{l(l+1)-2\beta q}\right)^{l+\frac{1}{2}}~,
\end{equation}
and $\mathcal{Z}_{\text{saddle}}[\Lambda] $ is defined in (\ref{ZLclas}). In (\ref{Zpert22}), we keep only quadratic terms in the action (including those stemming from the Fadeev-Popov determinant) while keeping the $\beta$ dependence exact. 
We now discuss the various pieces in (\ref{Zpert22}). The first parenthesis in (\ref{Zpert22}) comes from the zero-mode integral, defined by analytic continuation \cite{Polchinski:1988ua}  as
\begin{equation}\label{zeromodei}
\left(\frac{\upsilon\Lambda_{\mathrm{uv}}}{\pi}\right)^{\frac{1}{2}}\int {\text{d}\delta \varphi_{00}} \, e^{\frac{1}{2\pi}(1-\beta^2)\delta\varphi_{00}^2} = {\pm i}  \left(\frac{2\pi \upsilon \Lambda_{\mathrm{uv}}}{1-\beta^2}\right)^{\frac{1}{2}} ~,
\end{equation}
where the overall sign is ambiguous unless fixed by an additional principle. Integrating over the three $l=1$ modes, including the $\delta$-function from the Fadeev-Popov procedure, yields the second parenthesis in (\ref{Zpert22}). We note that in the prescription of Polchinski \cite{Polchinski:1988ua} the three $l=1$ modes (\ref{eigenvalues}) are also Wick rotated back,  which would lead to the real valued phase $(\pm i)^{1+3} =  1$. We must also consider the five $l=2$ modes separately, and these yield the third parenthesis in (\ref{Zpert22}). The infinite product in (\ref{Zpert22}) is divergent, but can be computed in a variety of ways. We consider the heat kernel method \cite{Vassilevich:2003xt}, which is covariant. Using a heat kernel regularisation scheme (see for example appendix C of \cite{Anninos:2020hfj})
\begin{multline}\label{heatkernel}
-\frac{1}{2} \sum_{l=3}^\infty ({2l+1} ) \log \left( \frac{l(l+1) -2\beta q}{4\pi\Lambda_{\mathrm{uv}} \upsilon} \right) =\int \frac{\dd t}{2t} \bigg[\frac{1+e^{-t}}{(1-e^{-t})}\frac{2 e^{-\frac{t}{2} + i \nu t}}{(1-e^{-t})} \cr
+e^{-(\frac{7}{2} + i\nu )t} \bigg(\frac{5-2e^{t} -2e^{2t} +2 e^{4t} +7 e^{5t}}{1-e^{-t}} -e^{4t}\frac{(7e^{t}-5)(1-e^{-(5-2i\nu)t})}{(1-e^{-t})^2}\bigg)\bigg]
\end{multline}
For the leading contribution $\nu = 3i/2$ we identify
\begin{equation}
\chi_{\Delta =2}(t)  = \frac{e^{-2t}}{(1-e^{-t})}~,
\end{equation}
as the Harish-Chandra character of the $SO(1,2)$ discrete series representation with $\Delta =2$ and quadratic Casimir $\Delta(1-\Delta) = -2$. 
We note that $SO(1,2)$ is the isometry group of Lorentzian dS$_2$. The contributions to the integrand that do not stem from $\chi_{\Delta =2}(t)$ exhibit subleading divergent structure at small $t$.  
Explicitly we obtain
\begin{multline}\label{heatkernel}
-\frac{1}{2} \sum_{l=3}^\infty ({2l+1} ) \log \left( \frac{l(l+1) -2\beta q}{4\pi\Lambda_{\mathrm{uv}} \upsilon} \right) = -\frac{107+12\nu^2}{12}\log\left(\frac{2e^{-\gamma_E}}{\varepsilon}\right)+\frac{2}{\varepsilon^2}+\nu^2\cr
+\left(\frac{1}{2}-\Delta_+\right)\zeta'(0,\Delta_+)+ \left(\frac{1}{2}-\Delta_-\right)\zeta'(0,\Delta_-)+\zeta'(-1,\Delta_+)+\zeta'(-1,\Delta_-)~\cr
+\frac{3}{2}\log \beta^2+\frac{5}{2}\log(2+\beta^2)+\frac{1}{2}\log (-1+\beta^2) + \frac{9}{2}\log 2~.
\end{multline}
where $\nu \equiv \sqrt{-2\beta q-1/4}$, $\Delta_{\pm}= 1/2\pm i\nu$, and $\zeta(a,z)$ denotes the Hurwitz $\zeta$-function. 
The heat kernel regularization amounts to
\begin{equation}\label{bessel}
-\frac{1}{2} \log \frac{{\color{black}\tilde{\lambda}}}{4\pi \upsilon\Lambda_{\mathrm{uv}}} = \int_0^\infty \frac{\dd\tau}{2\tau} e^{-\frac{\varepsilon^2}{4\tau}-{\color{black}\tilde{\lambda}}\tau} =  K_0\left(\sqrt{{\color{black}\tilde{\lambda}} } \varepsilon \right) \approx -\frac{1}{2}\log \frac{\varepsilon^2\, e^{2\gamma_E}{{\color{black}\tilde{\lambda}}}}{4}~,
\end{equation}
where $\varepsilon$ is a small parameter given by $\varepsilon = {e^{-\gamma_E}}/{\sqrt{\pi \upsilon \Lambda_{\mathrm{uv}}}}$, and $\gamma_E$ denotes the Euler-Mascheroni constant. The $1/\varepsilon^2$ divergence is local with respect to $\tilde{g}_{ij}$ and can be absorbed into an appropriate local counterterm built from $\tilde{g}_{ij}$. The parameter $\tilde{\lambda}= \upsilon\lambda \in
 \mathbb{R}^+$ denotes an eigenvalue.
Applying to (\ref{heatkernel}) the relations \cite{Adamchik}\footnote{These identities are to be understood as yielding a real valued analytic expression at small $\beta$.} 
\begin{equation}
\zeta'(0,z)= \text{log}\Gamma(z)-\frac{1}{2}\log(2\pi)~,\quad \zeta'(-1,z)= \zeta'(-1) -\log G(z+1)+z \, \text{log}\Gamma(z)~,
\end{equation} 
with $G(z)$ the Barnes $G$ function, $\zeta'(-1)=1/12-\log A$ with $A$ denoting Glaisher's constant, and $\text{log}\Gamma(z)$ the logGamma function, we finally arrive at the small $\beta$ expansion for (\ref{ZL2})
\begin{multline}\label{areaCFT}
\mathcal{Z}_{tL}[\Lambda] \approx
 \frac{\pm i }{\text{vol}_{SO(3)}} \, \mathrm{const}\,\times \,e^{-\frac{1}{\beta^2}- \frac{1}{\beta^2}\log\left(4\pi  \beta^2\right)} \upsilon^{\frac{c_L}{6}}\Lambda_{\mathrm{uv}}^{\frac{7}{6}- \beta^2} \Lambda^{-\frac{1}{\beta^2}+1} \\ \times \left(\frac{1}{\beta} - \left(\frac{1}{6}- \frac{5a_1\pi}{2a_0}+2\gamma_E+ \log 4\pi + \ldots \right)\beta + \ldots \right)~,
\end{multline}
where 
\begin{equation}\label{A0}
\mathrm{const} \equiv a_0\times \frac{6  \times 2^{1/3} \sqrt{3} }{ A^2  \pi ^{5/6}} e^{-25/12}~.
\end{equation}
We note that the $\upsilon$ dependence in (\ref{areaCFT}) is $\upsilon^{{c_L}/{6}}$, with 
\begin{equation}
c_L = 26-c_{\text{m}} = -\frac{6}{\beta^2} +13 -6\beta^2~.
\end{equation}
The dependence $\upsilon^{{c_L}/{6}}$ can be recognised to take the standard form (\ref{Sent}) of a two-dimensional conformal field theory on the two-sphere \cite{Zamolodchikov:2001dz}. This is consistent with the proposal that timelike Liouville theory can be viewed as a two-dimensional conformal field theory \cite{Ribault:2015sxa,Ikhlef:2015eua,Bautista:2019jau}. Moreover, we observe that $c_L -26 + c_{\text{m}} = 0$, a necessary condition to avoid any dependence on the redundancy (\ref{redundant}). 
%

\subsection{Two-loop contribution}

\begin{figure}[H]
\begin{center}
\begin{tikzpicture}[scale=.6]

\draw[black] (-2.5,0) --(-.5,0);
\draw (0.5,0) circle (1cm);
\draw (-3.5,0) circle (1cm);

\draw (6.5,0) circle (1cm);
\draw[black] (7.5,0) --(5.5,0);

\draw (12.5,0) circle (1cm);
\draw (14.5,0) circle (1cm);

\end{tikzpicture}
\end{center}
\caption{Double-tadpoles, melons and cactus diagrams.}
\label{fig:diagramsValpha}
\end{figure}
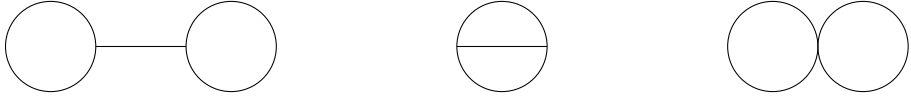

Beyond the Gaussian approximation, we must consider the contribution from higher loop Feynman diagrams. Given the perturbative action (\ref{Spert}), it follows that to leading order beyond the Gaussian approximation we must consider the contribution from the cubic and quartic vertex whose couplings go as $\sim \beta$ and $\sim \beta^2$ respectively. Given the constraints on specific angular momentum modes, it is convenient to express the computation in the space of spherical harmonics. 
As described in detail in appendix \ref{twoloop}, there are three connected diagrams that contribute. These are shown in figure \ref{fig:diagramsValpha}.  Two of the diagrams, a melonic diagram and a double-tadpole diagram, contain two cubic vertices. The third is a cactus diagram with a single quartic vertex.
\noindent
The double-tadpole diagram and the cactus diagram are logarithmically ultraviolet divergent. Nonetheless, due to the specific quartic and cubic couplings stemming from the Liouville action (\ref{Spert}),  the double-tadpoles and cactus diagrams exactly cancel each other  and consequently we do not have to deal with any ultraviolet divergences. We obtain the finite result
\begin{equation}\label{eq:diagrams}
\frac{1}{\mathcal{Z}^{(2)}_{\text{pert}}[\beta] } \times \mathcal{Z}_{\text{pert}}[\beta] = 1 + \mathlarger{\mathlarger{\ominus}}\,\beta^2 + \ldots~,
\end{equation}
where we defined 
%
%
%
%
%
\begin{equation}
\mathlarger{\mathlarger{\ominus}} \equiv -\frac{2}{3}\sum_{l_1,l_2,l_3\neq 1} \frac{(2l_1+1)(2l_2+1)(2l_3+1)}{(l_1(l_1+1)-2)(l_2(l_2+1)-2)(l_3(l_3+1)-2)} \begin{pmatrix} l_1 & l_2 & l_3 \\ 0 & 0 & 0\end{pmatrix}^2~,
\end{equation}
and
\begin{equation}
\begin{pmatrix} l_1 & l_2 & l_3 \\ m_1 & m_2 & m_3\end{pmatrix}~
\end{equation}
denotes the Wigner 3-j symbol. It is relatively straightforward to prove that $\mathlarger{\mathlarger{\mathlarger{\ominus}}}$ converges. The ellipses in (\ref{eq:diagrams}) indicate diagrams of order $\mathcal{O}(\beta^4)$. 


\subsection{Final two-loop expression \& all-loop expectations}

We can now assemble all the pieces and write down the expression up to two-loop contributions:
\begin{multline}\label{eq:SL_final_pathInt}
\mathcal{Z}_{tL}[\Lambda] \approx \frac{\pm i }{\text{vol}_{SO(3)}} \mathrm{const}\times e^{-\frac{1}{\beta^2}- \frac{1}{\beta^2}\log\left(4\pi  \beta^2\right)} \times \upsilon^{\frac{c_L}{6}}\Lambda_{\mathrm{uv}}^{\frac{7}{6}- \beta^2} \Lambda^{-\frac{1}{\beta^2}+1} \cr
\times \left(\frac{1}{\beta} - \left(\frac{1}{6}-\frac{5a_1\pi}{2a_0}+2\gamma_E+ \log 4\pi - \mathlarger{\mathlarger{\ominus}} \right)\beta+ \ldots \right)~.
\end{multline}
The ratio $a_1/a_0=+27/20\pi$\footnote{Note that the sign of $a_1/a_0$ has been corrected in v3. It is due to the Wick rotation of $\delta\varphi_{2,m}$ in the Fadeev-Popov determinant (\ref{FPexact}).} follows from (\ref{a0a1a2}), while the constant `const' is defined in (\ref{A0}). The ellipses in (\ref{eq:SL_final_pathInt}) indicate subleading terms in odd powers of $\beta$ starting at order $\mathcal{O}(\beta^3)$.\newline\newline
We end this section with some general observations and expectations about $\mathcal{Z}_{tL}[\Lambda]$:
\begin{itemize}
\item Although the perturbative $\beta$ expansion of  $\mathcal{Z}_{tL}[\Lambda]$ stems from a perturbative action (\ref{Spert}) which becomes complex along the pure imaginary $\delta\varphi$ contour, $\mathcal{Z}_{tL}[\Lambda]$ is pure imaginary to all orders in the small $\beta$ expansion. \
\item The Fadeev-Popov determinant $\Delta_{\text{FP}}$ (\ref{FPexact}) contributes a term at the same order as the two-loop contribution. 
\item The dependence on $\upsilon$ and $\Lambda$ in the pre-factor of (\ref{eq:SL_final_pathInt}) is exact and will receive no further corrections at subleading order.
\item The dependence of $\Lambda_{\text{uv}}$ is such that $\mathcal{Z}_{tL}[\Lambda]$ in (\ref{eq:SL_final_pathInt}) is dimensionless. At small $\beta$, the choice of $\Lambda_{\text{uv}}$ will affect the $\mathcal{O}(\beta)$ term in the parenthesis. Adding contributions from the matter and $\mathfrak{b}\mathfrak{c}$-ghost sectors will cancel the $\upsilon$ dependence but not the $\Lambda_{\text{uv}}$ dependence.

\item There will be no ultraviolet divergences at higher orders in the quantum loop expansion. This is often a property of supersymmetric theories. Here it is due to the specific values  of the interaction vertices in the perturbative action (\ref{Spert}).  In particular this indicates that if timelike Liouville theory is indeed a conformal theory, $\mathcal{O}_\beta = e^{2\beta \varphi}$ is an exactly marginal deformation. 

\item Conformal primaries of the matter theory with conformal weight $(\Delta,\bar{\Delta})$ can be consistently coupled to the gravitational theory provided they have $\Delta=\bar{\Delta}$ and are dressed by a Liouville primary of the form $\mathcal{O}_\alpha = e^{2\alpha \varphi}$ with $\alpha = \left( \sqrt{c_{\text{m}} -1 -24\Delta}- \sqrt{c_{\text{m}}-25} \right)/2\sqrt{6}$. We notice that $\Delta=\bar{\Delta}=1$ gives $\alpha=0$, while $\Delta\gg c_{\text{m}}$ gives $\alpha \approx - q/2 + i \sqrt{\Delta}$, with corrections only for the imaginary part \cite{Bautista:2020obj}. For $\Delta=\bar{\Delta} = 0$ we recover $\alpha = \beta$.  

\end{itemize}
It would be interesting to prove these expectations. In the next section we take a different route, and compare to an expression from the two-sphere partition function that is derived from the DOZZ formula \cite{Zamolodchikov:1995aa,Dorn:1994xn,Teschner:2001rv,Teschner:1995yf} for the three-point function applied to three exponential Liouville operators.

\section{Sphere-partition function from (timelike) DOZZ}\label{dozzsec}

Having established a systematic semiclassical expansion of timelike Liouville theory we can ask how our expressions compare to other methods in the literature. The natural object to compare to is derived from the expectation value of three exponential operators $\mathcal{O}_{\beta} = e^{2\beta\varphi}$. Assuming that timelike Liouville theory is conformally invariant at the quantum level \cite{Ribault:2015sxa,Ikhlef:2015eua,Bautista:2019jau}, as evidenced for example by (\ref{areaCFT}), the three-point function
\begin{equation}
\langle \mathcal{O}_{\beta}(z_1) \mathcal{O}_{\beta}(z_2) \mathcal{O}_{\beta}(z_3) \rangle =  \frac{1}{\text{vol}_{PSL(2,\mathbb{C})}} \times \frac{C(\beta,\beta,\beta)}{|z_1-z_2|^2 |z_1-z_3|^2 |z_2-z_3|^2}~,
\end{equation}
is fixed up to a coefficient $C(\beta,\beta,\beta)$. To make contact with the Liouville partition function $\mathcal{Z}_{tL
}[\Lambda]$ in (\ref{ZL}) we note that 
\begin{equation}
- \partial_\Lambda^3\mathcal{Z}_{tL}[\Lambda] = 2 \times C(\beta,\beta,\beta)~,
\end{equation}
where we have used that \cite{Zamolodchikov:1995aa}
\begin{equation}
\int_{\mathbb{C}^3} \frac{\dd^2 z_1 \dd^2 z_2 \dd^2 z_3}{|z_1-z_2|^2 |z_1-z_3|^2 |z_2-z_3|^2} =  2 \, {\text{vol}_{PSL(2,\mathbb{C})}}~.
\end{equation}
The integral over the three-points diverges logarithmically due to the regions of integration where two of the points collide. As is known from the critical string, these divergences are of the same magnitude as the divergent volume of $PSL(2,\mathbb{C})$ such that the latter can be used to fix the three-points and avoid performing the integral altogether. 

It follows from (\ref{eq:SL_final_pathInt}) that the small $\beta$ expansion of $C(\beta,\beta,\beta)$ about the real saddle (\ref{saddle}) is given by 
\begin{multline}\label{Cbbbpert}
C(\beta,\beta,\beta) \approx \frac{\pm i }{2\text{vol}_{SO(3)}} s^{\frac{7}{6}}\times \mathrm{const}\times e^{-\frac{1}{\beta^2}- \frac{1}{\beta^2}\log\left(4\pi  \beta^2\right) } \times \upsilon^{\frac{c_L}{6}}\tilde{\Lambda}_{\mathrm{uv}}^{\frac{7}{6}-\beta^2} \Lambda^{-\frac{1}{\beta^2}-2} \cr
\times \left(\frac{1}{\beta^7}  - \left(\frac{1}{6}- \frac{5a_1\pi}{2a_0}+2\gamma_E+ \log 4\pi  -\mathlarger{\mathlarger{\ominus}} {+} \log s \right)\frac{1}{\beta^5} + \ldots \right)~,
\end{multline}
where we have taken the ultraviolet scale $\tilde{\Lambda}_{\text{uv}}$ in the computation of the DOZZ formula to be linearly related to the ultraviolet scale $\Lambda_{\text{uv}}$ in the path-integral computation as ${\Lambda}_{\text{uv}} = s \tilde{\Lambda}_{\text{uv}}$. The $\beta$-independent dimensionless coefficient $s$ affects the $1/\beta^5$ coefficient. 

\subsection{Analytic continuation of spacelike Liouville?}

There have been several proposals for $C(\beta,\beta,\beta)$ in the literature. The proposal of \cite{Harlow:2011ny, Zamolodchikov:2005fy, Kostov:2005kk}  taken at face value leads to a vanishing result --- $C(\beta,\beta,\beta) = 0$. Here, instead, we would like to compare our result (\ref{Cbbbpert}) to the analytic continuation of the spacelike Liouville structure constant for the three-point function of $\mathcal{O}_b = e^{2b\varphi}$. This is a special case of the DOZZ formula \cite{Zamolodchikov:1995aa,Dorn:1994xn,Teschner:2001rv,Teschner:1995yf} and is given by
\begin{equation}\label{dozz}
\mathcal{C}(b,b,b) 
= -e^{-Q^2+ Q^2\log 4}\Lambda^{Q/b-3}(\pi\gamma(b^2))^{Q/b}\frac{(1-b^2)^2}{\pi^3b^5 \gamma(b^2)\gamma(b^{-2})}~,
\end{equation}
where $\gamma(z) \equiv \Gamma(z)/\Gamma(1-z)$ is a meromorphic function with poles when $z$ is a non-positive integer and zeros when $z$ is a non-vanishing positive integer.
At least semiclassically, the DOZZ formula can be computed from the Liouville path integral \cite{Harlow:2011ny}. To arrive at (\ref{dozz}), which differs slightly from the standard expression in the literature, we must take into account some additional normalisation factors. In the literature the path integral approach to obtaining the DOZZ formula involves first mapping the spacelike Liouville theory on the sphere to the disk using (\ref{phiT}). Concretely, upon transforming the spacelike Liouville action on the round two-sphere (\ref{FSmetric}) with $\upsilon=1$ to the disk using the field redefinition \cite{Harlow:2011ny}
\begin{equation}
\varphi \rightarrow \varphi - Q\log \frac{2}{1+z \bar{z}}~,
\end{equation}
a semiclassical comparison with (\ref{eq:SL_final_pathInt}), which is computed for the theory on the two-sphere, is only possible upon keeping the field-independent constants. Keeping all these terms, the spacelike Liouville action on a disk of radius $\kappa \gg 1$ reads
\begin{equation}
S_{L} [\varphi]= \frac{1}{4\pi}\int_d \dd r \dd \theta \sqrt{\tilde{g}_d} \left(\tilde{g}^{ab}_d \partial_a\varphi \partial_b \varphi + 4 \pi \Lambda\, e^{2b\varphi}\right)+ \frac{Q}{\pi}\oint_{\partial d}\dd \theta\varphi+ 2Q^2 \log \kappa + Q^2 - Q^2\log 4~,
\end{equation}
where $d\tilde{s}_d^2= \dd r^2+ r^2 \dd\theta^2$ with $r\in(0,\kappa)$. Any remaining terms vanish as we take $\kappa\to\infty$. Incorporating the field independent terms, leads to an additional factor in the DOZZ formula leading to the expression (\ref{dozz}).

The resulting expression (\ref{dozz}) allows  {\cite{Giribet:2011zx,McElgin:2007ak}  for a well defined analytic continuation $b\rightarrow \pm i\beta$, $Q\rightarrow  \mp i q$. The two-sphere partition function $\mathcal{Z}^{\text{DOZZ}}_{tL}[\Lambda]$, as derived from the DOZZ formula (\ref{dozz}) analytically continued to $\mathcal{C}(\beta,\beta,\beta)$, can be obtained by integrating $\mathcal{C}(\beta,\beta,\beta)$ three times with respect to $\Lambda$, and setting the integration constants to zero. The resulting expression --- along the lines of \cite{Giribet:2011zx} --- reads
\begin{equation}\label{tLZs2}
\mathcal{Z}^{\text{DOZZ}}_{tL}[\Lambda] = \pm i\left(\pi \Lambda \gamma(-\beta^2)\right)^{-\frac{1}{\beta^2}+1}\frac{(1+\beta^2)}{\pi^3 q \gamma(-\beta^2)\gamma(-\beta^{-2})}\,e^{q^2 -q^2 \log 4}~,
\end{equation}
where we recall that $q=1/\beta-\beta$. In a small positive $\beta$ expansion we find
\begin{equation}\label{eq:SL_final_DOZZ}
\mathcal{Z}^{\text{DOZZ}}_{tL}[\Lambda]  \approx \pm \frac{16}{\pi^2}e^{-2-2\gamma_E}\,\Lambda^{-\frac{1}{\beta^2}+1}\,\left( 1 - e^{\frac{2i\pi}{\beta^2}}\right)\times e^{-\frac{1}{\beta^2}- \frac{1}{\beta^2}\log (4\pi \beta^2)}\,\left(\frac{1}{\beta}+ {\left(\frac{19}{6}- \log 4\right)}\beta+ \ldots \right)\, ~,
\end{equation}
where we have taken $-1=e^{-i\pi}$. Given that the timelike Liouville action (\ref{eq:StL}) can be obtained by continuing $b$ to either positive or negative imaginary values, we interpret  there to be an overall sign ambiguity in (\ref{eq:SL_final_DOZZ}).

\subsection{Comparison}
At this stage we can compare the expressions (\ref{eq:SL_final_pathInt}) (with $\upsilon=1$) and (\ref{eq:SL_final_DOZZ}). For convenience we recall both expressions
\begin{eqnarray}\nonumber
\mathcal{Z}^{\text{DOZZ}}_{tL}[\Lambda]  &\approx& \pm  e^{-\frac{1}{\beta^2}- \frac{1}{\beta^2}\log (4\pi \beta^2) }\Lambda^{-\frac{1}{\beta^2}+1} \left( 1 - e^{\frac{2i\pi}{\beta^2}}\right)\,\left(\frac{1}{\beta}+ {\left(\frac{19}{6}- 2 \log 2\right)}\beta+ \ldots \right)~, \\ \nonumber
\mathcal{Z}_{tL}[\Lambda]  &\approx&  {\pm \, i } \, e^{-\frac{1}{\beta^2}- \frac{1}{\beta^2}\log\left(4\pi  \beta^2\right)} {\Lambda}^{-\frac{1}{\beta^2}+1} \tilde{\Lambda}_{\mathrm{uv}}^{\frac{7}{6}-\beta^2} 
 \left(\frac{1}{\beta} - \left(\frac{1}{6}- \frac{5a_1\pi}{2a_0}+2\gamma_E+ \log 4\pi - \mathlarger{\mathlarger{\ominus}} {+} \log s\right)\beta+ \ldots \right)~.
\end{eqnarray}
We recall that $s$ was implicitly defined in (\ref{Cbbbpert}), and we further note that we have rescaled both expressions such that the coefficient of the $1/\beta$ term in the parenthesis is one.
\begin{itemize}
\item The $\Lambda$ dependence between the two expressions agrees. In contrast, $\mathcal{Z}^{\text{DOZZ}}_{tL}[\Lambda]$ exhibits no $\Lambda_{\text{uv}} = s \tilde{\Lambda}_{\text{uv}}$ dependence. We interpret this to mean that for $\mathcal{Z}^{\text{DOZZ}}_{tL}[\Lambda]$, dimensionful quantities are measured in units of some choice of ultraviolet cutoff scale. Any discrepancy between the path integral and DOZZ based on choices of $\Lambda_{\text{uv}}$ is encoded in $s$. 
\item The DOZZ formula has an overall factor $\left( 1 - e^{{2i\pi}/{\beta^2}}\right)$. We interpret this to mean that $\mathcal{Z}^{\text{DOZZ}}_{tL}[\Lambda]$ receives contributions from two saddle point solutions of (\ref{eoms}). Particularly, the $e^{{2i\pi}/{\beta^2}}$ term can be viewed as stemming from a complex saddle $\varphi_* + \pi i /\beta$, where $\varphi_*$ is given by (\ref{saddle}). Complex saddles were studied in detail in \cite{Harlow:2011ny}, where it was argued that they indeed contribute to the Liouville path integral. It is interesting that $\mathcal{Z}^{\text{DOZZ}}_{tL}[\Lambda]$ seems to require only one additional saddle.
\item The leading order contribution to the small $\beta$ expansion multiplying the classical saddle point approximation is $1/\beta$ for both expressions. From the perspective of our computation this stems from the $\mathcal{O}(\beta^{-3})$ term in the Fadeev-Popov determinant (\ref{FPexact}). This term was anticipated in \cite{Harlow:2011ny}.
\item Both expressions have an overall sign ambiguity, but the real saddle contribution of $\mathcal{Z}^{\text{DOZZ}}_{tL}[\Lambda]$ differs from $\mathcal{Z}_{tL}[\Lambda]$ by an overall factor of $i$. It is worth noting here that the defining recursion relation \cite{Teschner:1995yf} satisfied by the DOZZ formula is ambiguous up to an overall rescaling, so in principle one could have defined the DOZZ formula with an overall $i$. From our perspective both this $i$ as well as the overall sign ambiguity are associated to the unsupressed zero mode (\ref{zeromodei}).
\item Picking $s = e^{-\frac{10}{3}+ \frac{5a_1}{2a_0}\pi-2\gamma_E+\ominus}/\pi$, we can match the $\mathcal{O}(\beta)$ term in the two expressions. This is the only remaining ambiguity (provided the measure over the Liouville field is indeed the flat measure \cite{DHoker:1990dfh,DHoker:1990prw}). Contributions from higher loop diagrams could be unambiguous and it would be very interesting to test against $\mathcal{Z}^{\text{DOZZ}}_{tL}[\Lambda]$ at $\mathcal{O}(\beta^3)$.  
\end{itemize}


\section{Outlook} \label{outlook}

Two-dimensional quantum gravity coupled to conformal matter with a large $c_{\text{m}}$ is arguably the simplest gravitational theory interacting with matter which moreover admits semi-classical de Sitter vacua. (For large and negative $c_{\text{m}}$ one also finds semiclassical Euclidean dS$_2$ solutions upon fixing the area \cite{Zamolodvarphikov:1982vx, talk_beatrix, Muhlmann:2021clm, db}.) As such, it constitutes a natural playground to test various proposals regarding quantum features for a de Sitter universe. In particular, one may hope for an exact expression of the Gibbons-Hawking path integral (\ref{zgrav0}) at least for vanishing genus. The main result of this paper, building on \cite{Zamolodvarphikov:1982vx,Harlow:2011ny}, is a concrete procedure to systematically compute this path integral in a semiclassical expansion. We have applied this procedure up to two-loops resulting in the expression (\ref{eq:SL_final_pathInt}). Combined with insights from the DOZZ formula and its analytic continuation to timelike Liouville theory, and recalling (\ref{g0Z}), we arrive at the following conjecture for the exact genus zero partition function of gravity coupled to a two-dimensional conformal field theory: 
\begin{equation}\label{conj}
\mathcal{Z}^{(0)}_{\text{grav}}[\Lambda]  = \mp   e^{2\vartheta}\, \mathcal{A} \, \left(\pi \Lambda \gamma(-\beta^2)\right)^{-\frac{1}{\beta^2} +1}\frac{(1+\beta^2)}{q \gamma(-\beta^2)\gamma(-\beta^{-2})}\,e^{q^2 -q^2 \log 4}~.
\end{equation}
We further recall that $q = 1/\beta-\beta$, and that $\beta$ is related to $c_{\text{m}}$ through (\ref{bexp}). 
In arriving at (\ref{conj}) we have multiplied (\ref{tLZs2}) by on overall factor of $i \mathcal{A}$, and we are working in units of $\tilde{\Lambda}_{\text{uv}}$. 

In the semiclassical limit $\beta \to 0^+$, 
the above expression can be interpreted as stemming from two saddles of $\varphi$, each accompanied by their respective perturbative series:
\begin{equation} 
\mathcal{Z}^{(0)}_{\text{grav}}[\Lambda]  \approx \pm  e^{2\vartheta}\, i \, \frac{\mathcal{A}}{\beta} \, \Lambda^{-\frac{1}{\beta^2} +1}  e^{-\frac{1}{\beta^2}- \frac{1}{\beta^2}\log (4\pi \beta^2) }  \left(\sum_{n=0}^\infty c_n \beta^{2n}  - e^{ \frac{2i\pi}{\beta^2}} \sum_{n=0}^\infty c_n \beta^{2n}   \right)~.
\end{equation}
The perturbative series about each saddle have the same coefficients up to an overall sign. This is because the saddles themselves are structurally similar, both constant and merely differing by an imaginary shift.  Normalising $c_0 = 1$, the first few coefficients are given by
\begin{equation}
c_1 = \frac{1}{6}(19-12 \log 2), \quad\quad c_2 = \frac{1}{288} \left(4 (19-12 \log 2)^2-192 \zeta (3)\right)~, \quad \ldots
 \end{equation} 
The real saddle corresponds to Euclidean dS$_2$ which is the round two-sphere. The contribution from this saddle is pure imaginary due to the overall $i$, a feature in common with odd-dimensional gravitational theories on a round sphere saddles \cite{Polchinski:1988ua}. The other saddle is a highly oscillatory complex function.  As a result of the complex saddle, $\mathcal{Z}^{(0)}_{\text{grav}}[\Lambda]$ is itself complex valued for generic $\beta$ rather than pure imaginary.\footnote{Given the wildly oscillatory nature of the complex saddle, it is amusing to consider integrating $\mathcal{Z}^{(0)}_{\text{grav}}[\Lambda]$ against a reasonably smooth distribution of $\beta$'s to remove it. This might resonate with some ideas expressed in \cite{Coleman:1988tj,Anninos:2013nra,Saad:2019lba,Maloney:2020nni,Afkhami-Jeddi:2020ezh}.} This contrasts its formal definition as a real valued path integral over the timelike Liouville action (\ref{eq:StL}) in an interesting way. For special values of $\beta$, the reality properties of $\mathcal{Z}^{(0)}_{\text{grav}}[\Lambda]$ simplify. We find that for $n\in\mathbb{Z}^+$, 
\begin{eqnarray}\label{discretebeta}
\beta = \sqrt{{1}/{n}}~  \quad\quad &\implies& \quad\quad  \mathcal{Z}^{(0)}_{\text{grav}}[\Lambda] = 0~, \\
\beta = \sqrt{{2}/(2n+1)}~  \quad\quad &\implies& \quad\quad  \mathcal{Z}^{(0)}_{\text{grav}}[\Lambda] \in i \, \mathbb{R}~.
\end{eqnarray}
A purely real $\mathcal{Z}^{(0)}_{\text{grav}}[\Lambda]$ requires a complex $\beta$, albeit one whose imaginary part can be made parametrically smaller than its real part. It would be interesting, and potentially resonate with other ideas in quantum gravity, if the consistency of our theory required $\beta$ to take discrete values (from a timelike Liouville theory perspective see also \cite{Ribault:2015sxa,Schomerus:2003vv}). A vanishing $\mathcal{Z}^{(0)}_{\text{grav}}[\Lambda]$ also follows form the timelike Liouville DOZZ formula proposed in \cite{Harlow:2011ny}, since it vanishes when evaluated for three $\mathcal{O}_\beta = e^{2\beta \varphi}$ operators. In particular we highlight that $\mathcal{Z}^{(0)}_{\text{grav}}[\Lambda]$ vanishes in the limit $\beta\rightarrow 1^{-}$ (\ref{discretebeta}), followed by $\Lambda \rightarrow 0^+$. One can view this limit as ending at the bosonic string with the 26 dimensional ($c_\text{m}=25$ plus $c_L=1$) flat target space. Our analysis therefore suggests a novel argument for the vanishing of the genus zero partition function of the critical bosonic string.

It is interesting to ask whether the real saddle accompanied by its infinite perturbative series expansion yields a function with finite radius of convergence. Somewhat remarkably, given that both saddles have precisely the same perturbative coefficients, we can divide (\ref{conj}) by $( 1 - e^{{2i\pi}/{\beta^2}})$ leading to the expression
\begin{equation}\label{conjii}
\tilde{\mathcal{Z}}^{(0)}_{\text{grav}}[\Lambda]  = \mp   e^{2\vartheta}\, \mathcal{A} \, {\left(\pi \Lambda \gamma(-\beta^2)\right)^{-\frac{1}{\beta^2} +1}} \frac{(1+\beta^2)}{q \gamma(-\beta^2)\gamma(-\beta^{-2})}\,{e^{q^2 -q^2 \log 4}}\times{{\left( 1 - e^{\frac{2i\pi}{\beta^2}}\right)}}^{-1}~.
\end{equation}
The small $\beta$ expansion of $\tilde{\mathcal{Z}}^{(0)}_{\text{grav}}[\Lambda]$ is given by that of the real Euclidean dS$_2$ saddle alone whose perturbative coefficients are $c_n$. It would be interesting to understand from a purely gravitational perspective whether we should include the additional saddle, as suggested from the DOZZ approach. 


In terms of $c_{\text{m}} > 25$, the $\Lambda$ dependence at genus zero is given by
\begin{equation}\label{fulllog}
\log \mathcal{Z}^{(0)}_{\text{grav}}[\Lambda]  = 2\vartheta + \frac{1}{12} \left(c_{\text{m}}+\sqrt{(c_{\text{m}}-25) (c_{\text{m}}-1)}-25\right) \log \frac{1}{\Lambda} + f^{(0)}(c_{\text{m}})~.
\end{equation}
This can be viewed as an all orders completion of (\ref{leadingZ}) in the $1/c_{\text{m}}$ expansion, where we understand the subleading terms as small corrections to the entanglement entropy across the dS$_2$ horizon due to the presence of dynamical gravity \cite{Anninos:2020geh}. For instance, the first correction to the $c_{\text{m}}/6$ coefficient in (\ref{leadingZ}) is $-19/6$. It would be interesting to test these ideas through other more direct methods \cite{Bombelli:1986rw,Srednicki:1993im}. According to the picture of Gibbons and Hawking \cite{Gibbons:1976ue,Gibbons:1977mu}, the additional terms  $\vartheta$ and $f^{(0)}(c_{\text{m}})$ encode further contributions to the dS$_2$ entropy.  Remarkably, an essentially equivalent formula to (\ref{fulllog}) holds for spacelike Liouville theory which is the theory one is led to when considering two-dimensional gravity coupled to a matter conformal field theory with $c_{\text{m}} < 1$. 

Finally, we comment on compact higher genus surfaces.\footnote{It is also worth considering manifolds with boundaries such as the disk partition function (see for example \cite{Saad:2019lba,Zamolodchikov:2001ah,Johnson:2019eik,Eberhardt:2021ynh}).} In even-dimensions the partition function of quantum field theory on a fixed background has a somewhat ambiguous overall normalisation (a discussion can be found in \cite{Gerchkovitz:2014gta}). For the specific case of two dimensions one can absorb the constant by shifting the coefficient $\vartheta$ of the Euler character viewed as a local counterterm. On the other hand, it should be emphasised that the detailed structure of $f^{(0)}(c_{\text{m}})$ encodes crucial information about the local field theoretic structure of the original theory. Further to this, the contribution from higher genus corrections renders the normalisation less ambiguous -- we can only fix the physical coupling $\vartheta$ for a single genus at most. For higher genus, based on the reasoning of \cite{Distler:1988jt,David:1988hj}, it is reasonable to anticipate the form
\begin{equation}
\log \mathcal{Z}^{(h)}_{\text{grav}}[\Lambda]  = \chi_h \vartheta  + \frac{\chi_h}{24} \left(c_{\text{m}}+\sqrt{(c_{\text{m}}-25) (c_{\text{m}}-1)}-25\right)  \log \frac{1}{\Lambda} + f^{(h)}(c_{\text{m}})~.
\end{equation}
If $\vartheta \gg 1$, these contributions will be exponentially small corrections as compared to $\log \mathcal{Z}^{(0)}_{\text{grav}}[\Lambda]$. It would be very interesting to compute $ f^{(h)}(c_{\text{m}})$ in the large $c_{\text{m}}$ limit for all $h$. Unlike the case of genus zero, the higher genus contributions will not be organised in terms of an expansion around a real saddle point solution. Perhaps this is remedied by fixing the area of the surface \cite{Zamolodvarphikov:1982vx,Seiberg:1990eb}.
%

\section*{Acknowledgements}
It is a pleasure to acknowledge Frederik Denef, Lorenz Eberhardt, Harold Erbin and Gaston Giribet for useful discussions. 
D.A. is funded by the Royal Society under the grant The Atoms of a deSitter Universe. The work of T.B. is supported by STFC grants ST/P000258/1 andST/T000759/1. The work of B.M. is part of the research programme of the Foundation for Fundamental Research on Matter (FOM), which is financially supported by the Netherlands Organisation for Science Research (NWO). 

\appendix

\section{Spherical harmonics}\label{Yapp}
We use real valued spherical harmonics throughout the paper. We denote by $\mathcal{Y}_{l m}(\theta,\phi)$ the complex spherical harmonics defined by 
\begin{equation}
\mathcal{Y}_{l m}(\theta,\phi)= \sqrt{\frac{(2l+1)}{4\pi}\frac{(l-m)!}{(l+m)!}}\, P_{l,m}(\cos\theta)\, e^{im\phi}~,
\end{equation}
where $P_{l,m}$ is the associated Legendre function, and $m \in [-l,l]$ with $l\in\mathbb{N}$. Real spherical harmonics $Y_{l m}(\theta,\phi)$ can be obtained using the linear combinations
\begin{equation}\label{real_complex}
{Y}_{l m}(\theta,\phi)= \begin{cases}
\frac{i}{\sqrt{2}}\left(\mathcal{Y}_{l m}(\theta,\phi)- (-1)^m \mathcal{Y}_{l, -m}(\theta,\phi)\right)~,\quad \mathrm{if}~ m<0\\
\mathcal{Y}_{l 0}(\theta,\phi)\\
\frac{1}{\sqrt{2}}\left(\mathcal{Y}_{l, -m}(\theta,\phi)+ (-1)^m \mathcal{Y}_{l m}(\theta,\phi)\right)~,\quad \mathrm{if}~ m>0~.
\end{cases}
\end{equation}
The Wigner 3-j symbol is given by the Clebsch-Gordan coefficients and gives the integral of the product of three complex spherical harmonics
\begin{align}
&\int \dd\phi \dd\theta\sin\theta\,\mathcal{Y}_{l_1,m_1}(\theta,\phi)\mathcal{Y}_{l_2,m_2}(\theta,\phi)\mathcal{Y}_{l_3,m_3}(\theta,\phi)\cr
&= \sqrt{\frac{(2l_1+1)(2l_2+1)(2l_3+1)}{4\pi}}\begin{pmatrix} l_1 & l_2 & l_3 \\ 0 & 0 & 0\end{pmatrix}\begin{pmatrix} l_1 & l_2 & l_3 \\ m_1 & m_2 & m_3\end{pmatrix}~.
\end{align}
\textbf{3-j symbol relations.}
The Clebsch-Gordan coefficients satisfy various properties. In particular they obey the orthogonality relation
\begin{equation}\label{orthoCG}
\sum_{\alpha,\beta}\begin{pmatrix} a & b & c \\ \alpha & \beta & \gamma\end{pmatrix}\begin{pmatrix} a & b & c' \\ \alpha & \beta & \gamma'\end{pmatrix}= \frac{1}{2c+1}\delta_{cc'}\delta_{\gamma\gamma'}~.
\end{equation}
Furthermore
\begin{equation}\label{reduced_CG}
\begin{pmatrix}
a & b & c \\ 
0 & 0 & 0 
\end{pmatrix} \neq 0\quad  \mathrm{iff} ~a+ b+ c \in 2\mathbb{Z}~\quad \&\quad \begin{pmatrix}
a & b & 0 \\ 
\alpha & \beta & 0 
\end{pmatrix}= \frac{(-1)^{a-\alpha}}{\sqrt{2a+1}}\delta_{ab}\delta_{\alpha-\beta}~.
\end{equation}
\section{Invariance under $PSL(2,\mathbb{C})$}\label{FPapp}
We explain the invariance of the Liouville action
\begin{equation}\label{eq:StL_app}
S_{tL}[\varphi] = \frac{1}{4\pi} \int \dd^2x\sqrt{\tilde{g}} \left( -\tilde{g}^{ij} \partial_i \varphi \partial_j \varphi -  q \tilde{R} \varphi + 4\pi \Lambda e^{2\beta\varphi}  \right)~,
\end{equation}
under (\ref{phiT})
\begin{equation}\label{phiT_app}
\varphi(z,\bar{z})\rightarrow \varphi(f(z),\overline{f(z)})+\frac{q}{2}\log\left(f'(z)\overline{f'(z)}\right)+q\left(\Omega(f(z),\overline{f}(z))-\Omega(z,\bar{z})\right)~.
\end{equation}
In the fiducial metric
\begin{equation}
\tilde{g}_{z\bar{z}}\dd z\dd\bar{z}\equiv e^{2\Omega(z,\bar{z})}\dd z\dd\bar{z}~.
\end{equation}
with Ricci scalar
\begin{equation}\label{Ricci_app}
\tilde{R}= -8\times e^{-2\Omega(z,\bar{z})}\partial_z\partial_{\bar{z}}\Omega(z,\bar{z})~.
\end{equation}
the Liouville action is given by
\begin{equation}\label{Liouville_app}
S_{tL}[\varphi]=\frac{1}{4\pi}\int \dd z\dd \bar{z}\left(2\varphi\partial_z\partial_{\bar{z}}\varphi-  q \sqrt{\tilde{g}}\, \tilde{R} \varphi+ 4\pi \Lambda \sqrt{\tilde{g}} e^{2\beta\varphi}\right)~.
\end{equation}
Now we show how each term individually transforms under (\ref{phiT_app}). We adopt the notation 
\begin{equation}
\varphi \equiv \varphi(z,\bar{z}), \quad \Omega\equiv \Omega(z,\bar{z})~,\quad \varphi_f \equiv \varphi(f(z),\overline{f(z)})~,\quad \Omega_f \equiv \Omega(f(z),\overline{f(z)})~.
\end{equation}
 For the last term it is quite easy, although importantly this term has to be invariant under  (\ref{phiT_app}) only in the semiclassical limit. We have 
\begin{align}
2\int \dd z\dd\bar{z} \sqrt{\tilde{g}}\Lambda e^{2\beta\varphi}&\rightarrow \int \dd z\dd\bar{z}e^{2\Omega}\left(f'(z)\overline{f'(z)}\right)\Lambda e^{2\left(\Omega_f-\Omega\right)}e^{2\beta\varphi_f}\cr
&=\int \dd f(z)\dd\overline{f(z)}e^{2\Omega_f}\Lambda e^{2\beta\varphi_f}=2\int \dd z\dd\bar{z} \sqrt{\tilde{g}}\Lambda e^{2\beta\varphi} ~.
\end{align}
The first term in the Liouville action (\ref{Liouville_app}) transforms under (\ref{phiT_app}) as
\begin{align}\label{kineticT}
&\int \dd z\dd \bar{z}\varphi\partial_z\partial_{\bar{z}}\varphi\cr
&\rightarrow \int \dd z\dd\bar{z}\varphi_f\partial_z\partial_{\bar{z}}\varphi_f+{2}q\int \dd z\dd\bar{z}\varphi_f\partial_z\partial_{\bar{z}}\left(\Omega_f-\Omega\right)  +q^2\int \dd z\dd\bar{z}\left(\Omega_f-\Omega\right)\partial_z\partial_{\bar{z}}\left(\Omega_f-\Omega \right)\cr
&=\int \dd z\dd \bar{z}\varphi\partial_z\partial_{\bar{z}}\varphi+q^2\int \dd z\dd\bar{z}\left(\Omega_f-\Omega\right)\partial_z\partial_{\bar{z}}\left(\Omega_f-\Omega\right)+{2}q\int \dd z\dd\bar{z}\varphi_f\partial_z\partial_{\bar{z}}\left(\Omega_f-\Omega\right) ~.
\end{align}
Finally combining the Liouville action  (\ref{Liouville_app}), the transformation (\ref{phiT_app}) and the Ricci scalar (\ref{Ricci_app}) we obtain for the term linear in $\varphi$
\begin{align}\label{linearT}
&\int \dd z\dd \bar{z}\sqrt{\tilde{g}}\,q \tilde{R} \varphi \rightarrow -{4}{q}\int \dd z\dd \bar{z}\left(\varphi_f+Q\left(\Omega_f-\Omega\right)\right)\partial_z\partial_{\bar{z}}\Omega\cr
&=-{4}q\int \dd z\dd \bar{z}\varphi_f\partial_z\partial_{\bar{z}}\Omega-{4}{q^2}\int \dd z\dd\bar{z}\left(\Omega_f-\Omega\right)\partial_z\partial_{\bar{z}}\Omega~.
\end{align}
Combining (\ref{kineticT}) and (\ref{linearT}) leads to
\begin{align}
&\int \dd z\dd \bar{z}\left(2\varphi\partial_z\partial_{\bar{z}}\varphi-\sqrt{\tilde{g}}\,q \tilde{R} \varphi\right) \rightarrow  2\int \dd z\dd \bar{z}\varphi\partial_z\partial_{\bar{z}}\varphi \cr
&+{4}{q}\int \dd z\dd\bar{z}\varphi_f\partial_z\partial_{\bar{z}}\Omega_f+{4}{q^2}\int \dd z\dd\bar{z}\left(\Omega_f-\Omega\right)\partial_z\partial_{\bar{z}}\Omega+2q^2\int \dd z\dd\bar{z}\left(\Omega_f-\Omega\right)\partial_z\partial_{\bar{z}}\left(\Omega_f-\Omega\right)\cr
&=2\int \dd z\dd \bar{z}\varphi\partial_z\partial_{\bar{z}}\varphi+{4}{q}\int\dd z\dd\bar{z}\varphi\partial_z\partial_{\bar{z}}\Omega+2q^2\int \dd z\dd\bar{z}\left(\Omega_f-\Omega\right)\partial_z\partial_{\bar{z}}\left(\Omega_f\right)+2q^2\int \dd z\dd\bar{z}\left(\Omega_f-\Omega\right)\partial_z\partial_{\bar{z}}\Omega\cr
&=2\int \dd z\dd \bar{z}\varphi\partial_z\partial_{\bar{z}}\varphi +{4}{q}\int\dd z\dd\bar{z}\varphi\partial_z\partial_{\bar{z}}\Omega+2q^2\int \dd z\dd\bar{z}\Omega_f\partial_z\partial_{\bar{z}}\Omega_f- 2q^2\int \dd z\dd\bar{z}\Omega\partial_z\partial_{\bar{z}}\Omega\cr
&+2q^2\int \dd z\dd\bar{z}\Omega_f\partial_z\partial_{\bar{z}}\Omega-2q^2\int \dd z\dd\bar{z}\Omega\partial_z\partial_{\bar{z}}\Omega_f
\cr&=\int \dd z\dd \bar{z}\left(2\varphi\partial_z\partial_{\bar{z}}\varphi-\sqrt{\tilde{g}}\,q \tilde{R} \varphi\right)~.
\end{align}
In summary, the Liouville action (\ref{Liouville_app}) is  invariant, semiclassically, under the transformation (\ref{phiT_app}).

\section{Fadeev-Popov gauge fixing}\label{FP}
We start with the Fubini-Study metric on the two-sphere and its transformation properties under $PSL(2,\mathbb{C})$. We have 
\begin{equation}
g_{ij}\dd x^i \dd x^j= e^{2b\varphi(z,\bar{z})}\tilde{g}_{z\bar{z}}\dd z \dd \bar{z}~, \quad \tilde{g}_{z\bar{z}}= \frac{4}{(1+ z\bar{z})^2}~.
\end{equation}
Now take $f(z)$ an \textit{arbitrary} diffeomorphism. The metric then transforms as 
\begin{equation}\label{eq:trafo_moebius}
d s^2= e^{2b\varphi(f(z),\overline{f(z)})+ \log f'(z)\overline{f'(z)}}\tilde{g}_{f(z)\overline{f(z)}}\dd z \dd \bar{z}~.
\end{equation}
Out of all the diffeomorphisms we consider those normalisable (\ref{norm_diffeos}) on the two-sphere. These are exactly the Moebius transformations
\begin{equation}
z\rightarrow f(z)= \frac{az+b}{cz+d}~, \quad ad-bc=1~, \quad a,b,c,d\in \mathbb{C}~.
\end{equation}
Analogously for $\overline{f(z)}$.
Out of the six real degrees of freedom the maximally compact $SO(3)$ subgroup of $PSL(2,\mathbb{C}$) restricts the parameters $a$, $b$, $c$, $d$ to  satisfy
\begin{equation}
b \bar{b}+ d \bar{d}=1~,\quad a\bar{b}+ c\bar{d}=0~,\quad a \bar{a}+ c \bar{c} =1~.
\end{equation}
These admit a parametrisation 
\begin{equation}
a=\cos\rho\, e^{i\alpha}~,\quad c= \sin\rho\,e^{i\gamma}~,\quad d= \bar{a}~,\quad b= -\bar{c}~,
\end{equation}
with $\alpha \in (0,2\pi)$, $\gamma \in (0,2\pi)$, and $\rho \in (0,\pi/2)$. Recalling that $a d - b c=1$ sets $\delta = 0$ such that we are left with three real parameters $\{\alpha,\rho,\gamma\}$ which are the Hopf coordinates of the $SO(3)$ group geometry which is the round three-sphere.
Infinitesimally, we find
\begin{equation}\label{deltaz}
\delta z= -\delta{c}z^2+(\delta{a}- \delta{d})z+\delta{b} 
\end{equation}
and 
\begin{equation}\label{eq:trafo_inf}
a=1+\delta{a}~, \quad b=0+\delta{b}~, \quad c=0+ \delta{c}~, \quad d= 1+ \delta{d}~.
\end{equation}
In particular in order to preserve the condition $ad-bc$ we must additionally impose that $\delta{a}= -\delta{d}$. The latter condition in particular implies that $\{\alpha,\rho,\gamma\}$ need to obey the constraints
\begin{equation}\label{eq:SO3}
{\color{black}\delta{a}=-\delta{d}=i\delta{\alpha}~, \quad \delta{b}=-\delta{\bar{c}}= \delta{\rho}e^{i\gamma}~,\quad \alpha=\rho=0~.}
\end{equation}
The above generate the three infinitesimal $SO(3)$ transformations.
\noindent
To gauge fix the residual $PSL(2,\mathbb{C})$ we are interested in those transformations in $PSL(2,\mathbb{C})/SO(3)$. Since $\{\alpha,\rho,\gamma\}$ are real valued (\ref{eq:SO3}) implies $\delta d\in \mathbb{R}$, as well as
\begin{equation}
\delta b+ \delta c=\delta\rho e^{i\gamma}-\delta \rho^*e^{-i\gamma}~,\quad \delta b- \delta c=\delta\rho e^{i\gamma}+\delta \rho^*e^{-i\gamma}~.
\end{equation}
Additionally we have
\begin{equation}
\delta b+ \delta c=2i \delta\rho \sin\gamma ~,\quad \delta b- \delta c=2\delta\rho \cos\gamma~,
\end{equation}
and consequently the condition $PSL(2,\mathbb{C})/SO(3)$ implies that $(\delta b+\delta c)\in \mathbb{R}$ and $(\delta b-\delta c)\in i\mathbb{R}$. In summary we have 
\begin{equation}\label{PSL2C}
{\color{black}\delta d\in \mathbb{R}~,\quad  (\delta b+ \delta c) \in \mathbb{R}~,\quad  (\delta b- \delta c) \in i\mathbb{R}~.}
\end{equation}
We now introduce spherical coordinates
\begin{equation}
z= e^{i\phi}\,\tan\frac{\theta}{2}\,~, \quad \bar{z}= e^{-i\phi} \tan\frac{\theta}{2}\, \quad \Rightarrow \quad \tilde{g}_{z\bar{z}}\dd z \dd \bar{z}= \dd \theta^2+ \sin^2\theta\dd \phi^2~,
\end{equation}
where $\theta \in (0,\pi)$ and $\phi \sim \phi+2\pi$.
The Liouville mode expanded in a basis of real spherical harmonics (\ref{real_complex}) then reads
\begin{equation}
\varphi(z,\bar{z})= \sum_{l, m}\varphi_{l m}{Y}_{l m}(\theta,\phi)~,
\end{equation}
where we are summing over $l\in [0,\infty)$ and $m\in [-l,l]$.
Under a Moebius transformation the transformation of the physical metric (\ref{eq:trafo_moebius}) is then captured by the transformation 
\begin{equation}
\varphi(z,\bar{z})= \sum_{l, m}\varphi_{l m}{Y}_{l m}(z,\bar{z})\rightarrow   \sum_{l , m}{\varphi}_{l m}{Y}_{l m}(f(z),\overline{f(z)})+q\log\left(\sqrt{f'(z)\overline{f'(z)}}\frac{(1+|z|^2)}{(1+|f(z)|^2)}\right)~.
\end{equation}
Near the identity, we have 
\begin{equation}
\frac{(1+ |z|^2)}{(1+|f|^2)} \approx (1+ |z|^2)\frac{1}{(1+|z|^2)}\left(1- \frac{z\delta\bar{z}+\bar{z}\delta z}{(1+|z|^2)}\right)=1- \frac{z\delta\bar{z}+\bar{z}\delta z}{(1+|z|^2)}~,
\end{equation}
and 
\begin{equation}
\sqrt{f'(z)\overline{f'(z)}}= (1-\delta d-\delta c z)(1-\delta \bar{d}-\delta \bar{c}\bar{z})=1- (\delta d+\delta\bar{d})-(\delta c\,z+\delta\bar{c}\,\bar{z})~,
\end{equation}
where $\delta z$ is given in (\ref{deltaz}). 
We then find for the argument of the logarithm 
\begin{equation}
\sqrt{f'(z)\overline{f'(z)}}\,\frac{(1+|z|^2)}{(1+|f(z)|^2)} \approx 1+\frac{1}{2}(\delta z'+\delta \bar{z}')- \frac{z\delta\bar{z}+\bar{z}\delta z}{(1+|z|^2)}~.
\end{equation}
So in total we have 
\begin{align}
\varphi(z,\bar{z})&\rightarrow   \sum_{l, m}\varphi_{l m}{Y}_{l m}(f(z),\overline{f(z)})+\frac{q}{2}(\delta z'+\delta \bar{z}')- q\frac{z\delta\bar{z}+\bar{z}\delta z}{(1+|z|^2)}~\cr
 &= \sum_{l, m}{\varphi}_{l m}{Y}_{l m}(z,\bar{z})+ \sum_{l, m}{\varphi}_{l m}\left(\delta z \partial_z+\delta{\bar{z}}\partial_{\bar{z}}\right){Y}_{l m}(z,\bar{z})+\frac{q}{2}(\delta z'+\delta \bar{z}')- q\frac{z\delta\bar{z}+\bar{z}\delta z}{(1+|z|^2)}~\cr
& \equiv \sum_{l , m}{\varphi}_{l m}{Y}_{l m}(z,\bar{z})+ \delta\varphi_{l,m}~,
\end{align}
where 
\begin{equation}
\partial_z= e^{-i\phi}\frac{(1+\cos\theta)}{2}\,\partial_\theta-i e^{-i\phi}\frac{(1+\cos\theta)}{2\sin\theta}\,\partial_\phi~,
\end{equation}
and similarly for $\partial_{\bar{z}}$. Focusing on the holomorphic part of $\delta \varphi_{l m}$ (the anti-holomorphic follows analogously) we obtain
\begin{multline}
\delta \varphi = \sum_{l, m}\varphi_{l m}\left(\delta b\, e^{-i\phi}- 2\delta d\,\tan\frac{\theta}{2}-\delta c\, e^{i\phi}\tan^2\frac{\theta}{2}\right)\times \left(\frac{(1+\cos\theta)}{2}\,\partial_\theta Y_{l m}-i \frac{(1+\cos\theta)}{2\sin\theta}\,\partial_\phi Y_{l m}\right)\cr
-q\left(\delta c\,e^{i\phi} \tan\frac{\theta}{2}+\delta d+\frac{1}{2}e^{i\phi}\sin\theta\left(\delta \bar{b}- 2\delta \bar{d}\,e^{-i\phi}\tan\frac{\theta}{2}-\delta \bar{c}\, e^{-2i\phi}\tan^2\frac{\theta}{2}\right)\right)~.
\end{multline}
In the above $\delta b$, $\delta d$, $\delta c$ $\in \mathbb{C}$.

Adding the holomorphic and antiholomorphic pieces we finally obtain 
\allowdisplaybreaks 
\begin{align}
\delta\varphi_{1,-1}&= -\im(\delta b+\delta c)\varphi_{1,0}+2\im(\delta d)\varphi_{1,1}+\frac{3}{\sqrt{5}}\re(\delta b+\delta c)\varphi_{2,-2}+\frac{6}{\sqrt{5}}\,\re(\delta d)\varphi_{2,-1}\cr
&-\sqrt{\frac{3}{5}}\,\im (\delta b-\delta c)\varphi_{2,0}-\frac{3}{\sqrt{5}}\im(\delta b-\delta c)\varphi_{2,2}-2q\sqrt{\frac{\pi}{3}}\,\im(\delta b- \delta c)~,\cr
\delta \varphi_{1,0}&=\im(\delta b+\delta c)\varphi_{1,-1}+\re(\delta b-\delta c)\varphi_{1,1}+\frac{3}{\sqrt{5}}\im(\delta b-\delta c)\varphi_{2,-1}+4\sqrt{\frac{3}{5}}\,\re(\delta d)\varphi_{2,0}\cr
&+\frac{3}{\sqrt{5}}\re(\delta b+\delta c)\varphi_{2,1}-4q\sqrt{\frac{\pi}{3}}\,\re(\delta d)~,\cr
\delta \varphi_{1,1}&=-2\im (\delta d)\varphi_{1,-1}-\re(\delta b-\delta c)\varphi_{1,0}+\frac{3}{\sqrt{5}}\im(\delta b-\delta c)\varphi_{2,-2}-\sqrt{\frac{3}{5}}\,\re(\delta b+\delta c)\varphi_{2,0}\cr
&+\frac{6}{\sqrt{5}}\re(\delta d)\varphi_{2,1}+\frac{3}{\sqrt{5}}\re(\delta b+\delta c)\varphi_{2,2}-2q\sqrt{\frac{\pi}{3}}\,\re(\delta b+ \delta c)~.
\end{align}
Using (\ref{PSL2C}) this reduces to
\allowdisplaybreaks
\begin{align}
\delta\varphi_{1,-1}&= \frac{3}{\sqrt{5}}\re(\delta b+\delta c)\varphi_{2,-2}+\frac{6}{\sqrt{5}}\,\re(\delta d)\varphi_{2,-1}-\sqrt{\frac{3}{5}}\,\im (\delta b-\delta c)\varphi_{2,0}-\frac{3}{\sqrt{5}}\im(\delta b-\delta c)\varphi_{2,2}\cr
&-2q\sqrt{\frac{\pi}{3}}\,\im(\delta b- \delta c)~,\cr
\delta \varphi_{1,0}&=+\frac{3}{\sqrt{5}}\im(\delta b-\delta c)\varphi_{2,-1}+4\sqrt{\frac{3}{5}}\,\re(\delta d)\varphi_{2,0}+\frac{3}{\sqrt{5}}\re(\delta b+\delta c)\varphi_{2,1}-4q\sqrt{\frac{\pi}{3}}\,\re(\delta d)~,\cr
\delta \varphi_{1,1}&=\frac{3}{\sqrt{5}}\im(\delta b-\delta c)\varphi_{2,-2}-\sqrt{\frac{3}{5}}\,\re(\delta b+\delta c)\varphi_{2,0}+\frac{6}{\sqrt{5}}\re(\delta d)\varphi_{2,1}+\frac{3}{\sqrt{5}}\re(\delta b+\delta c)\varphi_{2,2}\cr
&-2q\sqrt{\frac{\pi}{3}}\,\re(\delta b+ \delta c)~.
\end{align}
Crucially all the $l=1$ dependency drops out. Finally denoting by $\alpha_n$ the set of transformations in $PSL(2,\mathbb{C})/SO(3)$ (\ref{PSL2C}) we calculate the Fadeev-Popov determinant
\begin{align}
\Delta_{\text{FP}}&=\det\frac{\delta \varphi_{1,m}}{\delta \alpha_n}\bigg|_{m\in \{\pm 1,0\}}= \det \begin{pmatrix}
\frac{\delta \varphi_{1,-1}}{\delta \re(\delta d)} & \frac{\delta \varphi_{1,-1}}{\delta \re(\delta b+\delta c)} & \frac{\delta \varphi_{1,-1}}{\delta \im(\delta b-\delta c)}  \\
\frac{\delta \varphi_{1,0}}{\delta \re(\delta d)} & \frac{\delta \varphi_{1,0}}{\delta \re(\delta b+\delta c)} & \frac{\delta \varphi_{1,0}}{\delta \im(\delta b-\delta c)}  \\
\frac{\delta \varphi_{1,1}}{\delta \re(\delta d)} & \frac{\delta \varphi_{1,1}}{\delta \re(\delta b+\delta c)} & \frac{\delta \varphi_{1,1}}{\delta \im(\delta b-\delta c)}  
\end{pmatrix}\cr
&=\det \begin{pmatrix}
\frac{6}{\sqrt{5}}\varphi_{2,-1} & \frac{3}{\sqrt{5}}\varphi_{2,-2} & -\sqrt{\frac{3}{5}}\varphi_{2,0}-\frac{3}{\sqrt{5}}\varphi_{2,2}-2q\sqrt{\frac{\pi}{3}}\\
4\sqrt{\frac{3}{5}}\varphi_{2,0}-4q\sqrt{\frac{\pi}{3}} & \frac{3}{\sqrt{5}}\varphi_{2,1} & \frac{3}{\sqrt{5}}\varphi_{2,-1}\\
\frac{6}{\sqrt{5}}\varphi_{2,-1} & \frac{3}{\sqrt{5}}\varphi_{2,2}- \sqrt{\frac{3}{5}}\varphi_{2,0}-2q\sqrt{\frac{\pi}{3}} & \frac{3}{\sqrt{5}}\varphi_{2,-2} &  
\end{pmatrix}~\cr
&=a_0q^3+a_1q \sum_{m=-2}^2\varphi_{2,m}^2+a_2\Big(\varphi_{2,0}^3+\frac{3}{2}\varphi_{2,0}(\varphi_{2,1}^2+\varphi_{2,-1}^2)+\frac{3}{2}\sqrt{3}\varphi_{2,2}(\varphi_{2,1}^2-\varphi_{2,-1}^2)\cr
&+3\sqrt{3}\varphi_{2,1}\varphi_{2,-1}\varphi_{2,-2}-3\varphi_{2,0}(\varphi_{2,-2}^2+\varphi_{2,2}^2)\Big)~,
\end{align}
where 
\begin{equation}
a_0\equiv -\frac{16}{3\sqrt{3}}\pi^{3/2}\,~,\quad a_1\equiv \frac{12}{5}\sqrt{3\pi}\,~,\quad a_2\equiv \frac{12}{5}\sqrt{\frac{3}{5}}~.
\end{equation}
In particular the Fadeev-Popov determinant $\Delta_{\text{FP}}$ is manifestly $SO(3)$ invariant. 

\section{Two-loop contribution}\label{twoloop}

In this appendix we calculate the two-loop corrections to the path integral (\ref{Zpert}):
\begin{align}\label{Zpert_app}
\mathcal{Z}_{\mathrm{pert}}[\beta]= \int [{\mathcal{D}'\delta\varphi}]\, e^{-S_{\mathrm{pert}}^{(2)}[\delta\varphi]}\,e^{-\frac{1}{4\pi} \frac{q}{\beta}\int\dd{\Omega}\left(-i\frac{4}{3}\beta^3 \delta\varphi(\Omega)^3+ \frac{2}{3}\beta^4\delta\varphi(\Omega)^4+\ldots \right)}~,
\end{align}
where $\Omega$ is a point on $S^2$ and $\dd \Omega$ denotes the volume element on the two-sphere. The prime on the measure indicates that we are removing the three $l=1$ modes.
First we note the propagator. For $\Omega, \Omega'$ two points on the round two-sphere we have
\begin{equation}\label{propagator}
G(\Omega;\Omega') \equiv  \frac{1}{\mathcal{Z}_{\mathrm{pert}}^{(2)}[\beta]} \int [\mathcal{D}'\delta\varphi] e^{-S_{\mathrm{pert}}^{(2)}[\delta\varphi]} \delta\varphi(\Omega) \delta\varphi(\Omega') = 2\pi \sum_{l\neq 1 ,m\in[-l,l]} \frac{{Y}_{l m}(\Omega) {Y}_{l m}(\Omega')}{A_l} ~,
\end{equation}
where we defined for $l\neq 1$, $A_l\equiv (l(l+1)-2+2\beta^2)$.
In particular at coincidence where $\Omega=\Omega'$ we have 
\begin{equation}\label{divloop}
\int\dd\Omega G(\Omega;\Omega) = {2\pi} \sum_{l\neq 1}^\infty \frac{2l+1}{l(l+1)-2+2\beta^2} = 4\pi G(\Omega_0;\Omega_0)~.
\end{equation}
The last equality follows from the fact that $G(\Omega;\Omega)$ is $\Omega$ independent. The above sum diverges logarithmically, as expected for coincident fields in two-dimensions. (\ref{divloop}) holds true also if we remove the $l=1$ modes since for each $l$ the spherical harmonics form an irreducible representation of $SO(3)$.

Now we expand the exponential up to order $\mathcal{O}(\beta^2)$ we obtain
\begin{multline}
\frac{1}{\mathcal{Z}_{\mathrm{pert}}^{(2)}[\beta]}\times \mathcal{Z}_{\mathrm{pert}}[\beta]\cr
=\frac{1}{\mathcal{Z}_{\mathrm{pert}}^{(2)}[\beta]}\int [{\mathcal{D}'\delta\varphi}]\, e^{-S_{\mathrm{pert}}^{(2)}[\delta\varphi]}\, \left(1-\frac{1}{2!} \frac{1}{9\pi^2}\beta^2 \int \dd \Omega\dd \Omega'\delta\varphi(\Omega)^3\varphi(\Omega')^3- \frac{1}{6\pi}\beta^2 \int \dd\Omega \delta\varphi(\Omega)^4+ \ldots\right)~,
\end{multline}
where $\mathcal{Z}_{\mathrm{pert}}^{(2)}[\beta]$ is the Gaussian part of $\mathcal{Z}_{\mathrm{pert}}[\beta]$ (\ref{Zpert22}). 
Performing Gaussian integrals we obtain
\begin{multline}\label{Zpert_app}
\frac{1}{\mathcal{Z}_{\mathrm{pert}}^{(2)}[\beta]}\times \mathcal{Z}_{\mathrm{pert}}[\beta]=1- \frac{1}{2!}\times \frac{1}{9\pi^2}\beta^2 c_1\int \dd\Omega \dd\Omega' G(\Omega,\Omega')^3 - \frac{1}{6\pi}\beta^2 c_3 \int \dd\Omega G(\Omega,\Omega)^2\cr- \frac{1}{2!}\times \frac{1}{9\pi^2}\beta^2c_2\int \dd\Omega \dd\Omega' G(\Omega,\Omega)G(\Omega,\Omega')G(\Omega',\Omega')+ \ldots~.
\end{multline}
In the above expression $c_1=6$, $c_2=9$ and $c_3=3$ are combinatorial expressions arising upon performing Gaussian integrals. \newline\newline
\textbf{Melonic term.} Although position space is useful to assess the nature of the ultraviolet divergences, it is not so useful to evaluate the diagrams. It is convenient to go to momentum space for this purpose. To do this, we can substitute the expression (\ref{propagator}) for $G(\Omega;\Omega')$ into the first term of (\ref{Zpert_app}) to obtain 
\begin{equation}\label{melon}
\mathlarger{\mathlarger{\ominus}} \equiv  -\frac{1}{18\pi^2} \, c_1\, (2\pi)^3\beta^2\sum_{\bold{l}\neq \textbf{1},\bold{m}}  \, \frac{W_{\bold{l},\bold{m}} W_{\bold{l},\bold{m}}}{A_{l_1} A_{l_2} A_{l_3}}~. 
\end{equation}
where $\bold{l} = (l_1,l_2,l_3)$, $\bold{m} = (m_1,m_2,m_3)$, and
\begin{equation}\label{eq:real3j}
W_{\bold{l},\bold{m}} \equiv \int\dd\Omega \, Y_{l_1 m_1}(\Omega)Y_{l_2 m_2}(\Omega)Y_{l_3 m_3}(\Omega)~,
\end{equation}
is a type of Wigner 3-j symbol for real spherical harmonics.
To evaluate (\ref{eq:real3j}) we replace real spherical harmonics by their complex counterpart using (\ref{real_complex}).  
If we expand in terms of complex spherical harmonics, we find
\begin{equation}\label{melonii}
\mathlarger{\mathlarger{\ominus}}=   - \frac{1}{18\pi^2} c_1\times  (2\pi)^3\beta^2\sum_{\bold{l}\neq \textbf{1},\bold{m}} \frac{\mathcal{W}_{\bold{l},\bold{m}} \mathcal{W}^*_{\bold{l},\bold{m}}}{ \, A_{l_1} A_{l_2} A_{l_3}}~,
\end{equation}
where $\bold{l} = (l_1,l_2,l_3)$, $\bold{m} = (m_1,m_2,m_3)$, and
\begin{eqnarray}
\mathcal{W}_{\bold{l},\bold{m}} &\equiv&  \int\dd\Omega \, \mathcal{Y}_{l_1 m_1}(\Omega)\mathcal{Y}_{l_2 m_2}(\Omega)\mathcal{Y}_{l_3 m_3}(\Omega) \\ 
&=& \sqrt{\frac{(2l_1+1)(2l_2+1)(2l_3+1)}{4\pi}} 
\begin{pmatrix}
l_1 & l_2 & l_3\\
0 & 0 & 0
\end{pmatrix}
\begin{pmatrix}
l_1 & l_2 & l_3\\
m_1 & m_2 & m_3
\end{pmatrix}~, 
\end{eqnarray}
where in the second line we have written it in terms of Wigner 3-j symbols. $\mathcal{W}^*_{\bold{l},\bold{m}}$ is the complex conjugate. Combining the orthogonality relation (\ref{orthoCG}) with the fact that each of the $m_i'$s itself run over $2l_i+1$ integers we obtain
\begin{equation}\label{melons_app}
\mathlarger{\mathlarger{\ominus}} = -\frac{2}{3} \beta^2\sum_{\bold{l}\neq \textbf{1}} \frac{(2l_1+1)(2l_2+1)(2l_3+1)}{A_{l_1}A_{l_2} A_{l_3}} \begin{pmatrix} l_1 & l_2 & l_3 \\ 0 & 0 & 0\end{pmatrix}^2~.
 \end{equation}~
\newline
{\textbf{Double-tadpole.}} A different type of contribution at order $\mathcal{O}(\beta^2)$ comes from contractions of the double-tadpole type
\begin{equation}\label{tp_beginning}
\bigcirc\!\!-\!\bigcirc= - \frac{1}{18\pi^2}c_2 \times \beta^2\times G(\Omega_0,\Omega_0)^2 \int\dd{\Omega\dd\Omega'}\, G(\Omega,\Omega')~,
\end{equation}
where $c_2=9$ is another combinatorial factor. By the orthogonality of the spherical harmonics the final integral 
is proportional to the $l=0$ mode. Consequently we have 
\begin{equation}\label{tp_intermediate}
\bigcirc\!\!-\!\!\bigcirc= - \frac{1}{18\pi^2}c_2 \times \frac{8\pi^2}{2^2}\times \beta^2\frac{1}{ A_0}\sum_{l, l'}\frac{(2l+1)(2l'+1)}{A_l A_{l'}} =-\frac{\beta^2}{A_0}\sum_{l,l'\neq 1}\frac{(2l+1)(2l'+1)}{A_l A_{l'}}
\end{equation}
This contribution diverges like the square of the logarithm.\newline\newline
\textbf{Cacti.} Performing Gaussian integrals for the quartic term in (\ref{Zpert_app}) we obtain what we will denote as cactus diagrams
\begin{equation}\label{cactus1}
\bigcirc\!\bigcirc = -  \frac{1}{6\pi}\beta^2 c_3 \int\dd\Omega \,G(\Omega,\Omega)^2= -4\pi\times \frac{1}{6\pi}c_3\beta^2  G(\Omega_0,\Omega_0)^2~,
\end{equation}
where $c_3 = 3$. Using (\ref{divloop}) we obtain for (\ref{cactus1})
\begin{equation}\label{cactus}
\bigcirc\!\bigcirc = - \frac{\beta^2}{2}\sum_{l,l'\neq 1}\frac{(2l+1)(2l'+1)}{A_l A_{l'}}~.
\end{equation}
This contribution diverges like the square of the logarithm.
\newline\newline
\textbf{Melons + double-tadpoles + cacti.} At order $\mathcal{O}(\beta^2)$ the contributions of (\ref{Zpert_app}) split up in three different types of diagrams: melons $\mathlarger{\mathlarger{\ominus}}$ (\ref{melons_app}), double-tadpoles ${\bigcirc\!\!-\!\!\bigcirc}$ (\ref{tp_intermediate}) and cactus diagrams ${\bigcirc\!\bigcirc}$ (\ref{cactus}). Since $A_0=-2$ the double-tadpoles and the cactus diagrams, which were responsible for the ultraviolet divergences, exactly cancel. In particular, had the relation between $q$ and $\beta$ been different such ultraviolet divergences would appear. We are thus left with
\begin{equation}
\frac{1}{\mathcal{Z}^{(2)}_{\text{pert}}[\beta] } \times \mathcal{Z}_{\text{pert}}[\beta] =1+\left(\mathlarger{\mathlarger{\ominus}}+ {\bigcirc\!\!-\!\!\bigcirc}+{\bigcirc\!\bigcirc}\right) \beta^2+ \ldots=1+ \mathlarger{\mathlarger{\ominus}}\, \beta^2+ \ldots ~,
\end{equation}
where the ellipses indicate diagrams of order $\mathcal{O}(\beta^4)$.

 \begingroup

\end{document}